\documentclass{emulateapj} \usepackage{apjfonts}
\usepackage{amstext}
\usepackage{amsmath}
\usepackage{mathletters}

\begin{document}
\submitted {{\apj}, in press}
\journalinfo {{\apj}, in press}

\shorttitle{THE $\sigma$ PROBLEM IN PULSAR WINDS \& MHD ASYMPTOTIC ANALYSIS}
\shortauthors{{VLAHAKIS}}

\title{Ideal MHD solution to the $\sigma$ problem in Crab-like pulsar winds
\\
and General Asymptotic Analysis of Magnetized Outflows}

\author{Nektarios Vlahakis}
\affil{Department of Astronomy \& Astrophysics and Enrico Fermi
Institute, University of Chicago, 5640 S. Ellis Ave., Chicago, IL 60637
\\ Present address:
Section of Astrophysics, Astronomy \& Mechanics,
Department of Physics, University of Athens, 15784 Zografos Athens, Greece \\
vlahakis@phys.uoa.gr}

\begin{abstract}
Using relativistic, steady, axisymmetric, ideal magnetohydrodynamics (MHD) we analyze the 
super-Alfv\'enic regime of a pulsar wind by means of solving the momentum equation
along the flow as well as in the transfield direction. Employing a self-similar model,
we demonstrate that ideal MHD can account for the full acceleration from
high ($\gg 1$) to low ($\ll 1$) values of $\sigma$, the Poynting-to-matter energy flux ratio.
The solutions also show a transition from a current-carrying
to a return-current regime, partly satisfying the current-closure condition.
We discuss the kind of the boundary conditions near the base of the ideal MHD regime
that are necessary in order to have the required transition from high to low $\sigma$
in realistic distances, and argue that this is a likely case for an equatorial wind. 
Examining the MHD asymptotics in general, we extend the analysis of
Heyvaerts \& Norman and Chiueh, Li, \& Begelman by including two new elements:
classes of quasi-conical and parabolic field line shapes
that do not preclude an efficient and much faster than logarithmic acceleration,
and the transition $\sigma=\sigma_{c}$ after which the centrifugal forces (poloidal and 
azimuthal) are the dominant terms in the transfield force-balance equation.
\end{abstract}


\section{Introduction}
\label{introduction}

The spin down luminosity of a rotating pulsar powers the emission of 
an associated synchrotron nebula
by driving an initially Poynting flux-dominated wind 
(the Poynting-to-matter energy flux ratio is $\sigma \gg 1$)
that flows along the open magnetic field lines.
The density of the outflowing matter is enhanced by pair-creating cascades near the source
and the resulting number density is sufficiently large to screen the electric field along 
the flow and provide the necessary charge and current densities for ideal magnetohydrodynamics
(MHD) to hold.

At a distance $r_{s} \approx 3 \times 10^{17}$cm 
--specifically for the Crab --
the ram pressure of the
wind equals the total nebula pressure and the wind terminates 
in a standing shock \citep{RG74}.
The ``wisps'' (which appear at distance $\sim r_{s}$ from the center)
are an observational signature of the coupling between 
the fast pulsar wind and the slowly expanding nebula \citep{H95}.
By applying ideal MHD in the nebula, the matching shock conditions
require that the pulsar wind has become completely matter-dominated at the
position of the shock; the $\sigma$ function should be as low as $\sigma_{s}= 0.003$ 
just upstream of the termination shock (Kennel \& Coroniti 1984a,b),
and the question arises as to how this transition from high to low values of $\sigma$
can happen, the so-called $\sigma$ problem.

We note, however, that the very existence of the shock has been questioned by
\citet{B02}, who argues that no shock is necessary and that the observed emission
is created by current dissipation in a Poynting flux-dominated wind.
{}Furthermore, even if a shock indeed terminates the pulsar wind, abandoning
the axisymmetry or the ideal MHD in the nebula results in 
undetermined $\sigma_{s}$ \citep{B98,KG02}.
Even so, $\sigma_{s}$ could in general be small, and the interpretation of the
wisps as ion driven compressions \citep{A02}, as well as the
fitting of the observed nebula emission 
\citep{BK02}, provide additional support for the value $\sigma_{s}=0.003$.

The Crab-pulsar/nebula is the first and most analyzed object of this kind.
Recent X-ray observations have shown that there exist other
objects with common characteristics [e.g., the Vela pulsar
\citep{HGH01}, the nebulae associated with PSR B1509-58 \citep{G02}, and
PSR B1957+20 \citep{S03}].

The study of approximately monopole magnetic field geometries that are extremely
inefficient accelerators gave the
impression to the community that ideal MHD is in general unable to solve the $\sigma$ problem,
and many authors refer to the ``$\sigma$ parameter'' and not to the 
``$\sigma$ {\emph{function}}''.
Moreover, several alternative (non-ideal MHD) attempts
(see, e.g., \citealp{A98} for a review) gave unsatisfactory results,
and the puzzle remains unsolved for almost three decades.
However, there are in the literature
$r$ self-similar, ideal MHD solutions that start with $\sigma \gg 1$ near the origin of the
flow and reach $\sigma_\infty \approx 1$ (\citealp*{LCB92}; \citealp{VK03a}),
or even more efficient accelerators with $\sigma_\infty \approx 0.5$ \citep{VK03b},
or $\sigma_\infty \approx 0.1$ \citep*{VPK03}.
(Actually, $r$ self-similar are the only for the present
known {\emph {exact}} solutions of 
relativistic MHD in the sense that they satisfy the momentum
equation in both directions: along the flow {\emph {and}} in the transfield direction.)
We note in this connection that the main acceleration mechanism is not
the centrifugal, but the {\emph{magnetic}}.
The former may be important in the sub-Alfv\'enic regime of a disk-wind and results in
poloidal velocity of the order of the initial Keplerian speed, while the latter --
caused by the magnetic pressure-gradient force -- results in much higher velocity
that depends on the initial Poynting-to-mass flux ratio $\mu c^2$.
Employing the $\sigma$ function we may write $\gamma_\infty = \mu / (1+\sigma_\infty)$,
thus the maximum final Lorentz factor is attained in solutions with $\sigma_\infty=0$.
In this paper we show that ideal MHD can account for the full acceleration
from $\sigma\gg 1$ to $\sigma \approx 0$,
{\emph {if}} the boundary condition $| I | >> A \Omega $ at the base
of the ideal MHD accelerating regime is satisfied (here 
$I$ is the poloidal current, $A$ the poloidal magnetic flux function, and
$\Omega$ the field angular velocity).

The transfield force-balance equation is the most important in determining the 
acceleration by means of controlling the field line shape and consequently
how fast the quantity $\varpi^2 B_p$ decreases 
(here $\varpi$ is the cylindrical distance and $B_p$ the poloidal magnetic field).
In the main part of the acceleration the transfield component of the electromagnetic
force equals the poloidal centrifugal force that is proportional to the
curvature of the poloidal field lines $1/ {\cal R}$.
If this is still the case after the end of the acceleration -- in the far-asymptotic regime 
where $\varpi / {\cal R} =0 $ -- the electromagnetic force exactly vanishes
yielding the solvability condition at infinity (\citealp{HN89}; \citealp*{CLB91}).
However, if during the acceleration phase sufficiently small $\sigma$ values
are reached, the azimuthal centrifugal force -- the other part of the inertial force --
becomes important.
In this regime it is the difference between poloidal and azimuthal centrifugal
forces that equals the much smaller electromagnetic force in the transfield direction,
and no solvability condition can be derived.

We organize the paper as follows.
In \S \ref{MHD} we present the ideal MHD formalism
focusing on the super-Alfv\'enic asymptotic regime
where the puzzling transition from $\sigma \gg 1 $ to $\sigma \ll 1$ happens.
In \S \ref{zss} we derive a novel $z$ self-similar class of solutions
describing the polar region of the wind and present the results of the integration.
In \S \ref{genanalysis} we analyze the asymptotic shape of the flow
in relation to the acceleration, demonstrating that the same results are
expected in non-self-similar cases as well.
In \S \ref{equatorialwind} we discuss the problem of the pulsar magnetosphere,
pointing out the problems that the force-free picture faces, and suggesting
qualitatively a scenario that seems plausible for creating an equatorial wind
with the appropriate conditions for efficient acceleration from high to low-$\sigma$ values
satisfied. A summary follows in \S \ref{conclusions}.

\section{The steady - axisymmetric MHD description}\label{MHD}
\subsection{Governing equations}\label{governing}
The system of equations of special relativistic, steady, ideal MHD,
consists of the Maxwell equations,
Ohm's law, and the continuity, entropy and momentum equations:
\begin{mathletters}
\begin{equation} \label{max1}
{\boldsymbol{\nabla}} \cdot {\boldsymbol{B}}=0 \,, \quad
{\boldsymbol{\nabla}} \times {\boldsymbol{E}}= 0\,,
\end{equation}
\begin{equation} \label{max2}
{\boldsymbol{\nabla}} \times {\boldsymbol{B}} = 4 \pi {\boldsymbol{J}} / c \,,
\quad 
{\boldsymbol{\nabla}} \cdot {\boldsymbol{E}} = 4 \pi J^0 / c \,,
\quad {\boldsymbol{E}}= {\boldsymbol{B}} \times {\boldsymbol{V}} / c \,,
\end{equation}
\begin{equation}\label{contin_energy}
{\boldsymbol{\nabla}} \cdot \left(\rho_0 \gamma {\boldsymbol{V}} \right)=0\,, \quad
{\boldsymbol{V}} \cdot {\boldsymbol{\nabla}} \left({P}/{\rho_0^{\Gamma}}\right)=0\,,
\end{equation}
\begin{equation}\label{momentum}
-\gamma \rho_0 \left({\boldsymbol{V}}\cdot {\boldsymbol{\nabla}} \right) 
\left(\xi \gamma {\boldsymbol{V}}\right)
-{\boldsymbol{\nabla}}P + 
\left(J^0 {\boldsymbol{E}} +{\boldsymbol{J}} \times {\boldsymbol{B}} \right) / c =0 \,.
\end{equation}
\end{mathletters}
\noindent
Here $\boldsymbol V$ is the velocity of the outflow,
$\gamma$ the associated Lorentz factor,
${\boldsymbol E}\,,{\boldsymbol B}$ 
the electric and magnetic fields as measured in the central object's frame,
$J^0/c \,,{\boldsymbol J}$ the charge and current densities,
$\rho_0\,,P$ the gas rest-mass density and pressure in the comoving frame,
while for the polytropic equation of state (with index $\Gamma$) 
the enthalpy-to-rest-mass ratio is
\begin{equation}\label{xi}
\xi c^2 = c^2 + \frac{\Gamma}{\Gamma-1} \frac{P}{\rho_0 } \,.
\end{equation}

Assuming axisymmetry [$\partial/\partial \phi=0$,
in spherical $\left(r\,,\theta\,,\phi\right)$ and
cylindrical $\left(z\,,\varpi\,,\phi\right)$ coordinates
with $\hat{z}$ along the rotation axis and the central object
at ($\varpi=0\,, z=z_{c})$],
five conserved quantities along the flow exist.
If $A$ is the poloidal magnetic flux function,
they are (e.g., \citealt{VK03a}):
\begin{mathletters}
\begin{eqnarray} 
\label{Psi}
& 
& 
\hspace{-.7cm}
\mbox{the mass-to-magnetic flux ratio}
\
\Psi_A(A)=\frac{4 \pi \gamma \rho_0 V_p}{B_p}\,,
\\ 
\label{omega}
& 
& 
\hspace{-.7cm}
\mbox{the field angular velocity}
\
\Omega(A)= \frac{V_\phi}{\varpi}-\frac{V_p}{\varpi} \frac{B_\phi}{B_p}\,,
\\ 
\label{L}
& 
& 
\hspace{-.7cm}
\mbox{the specific angular momentum}
\
L(A)=\xi \gamma \varpi V_\phi -\frac{\varpi B_\phi}{\Psi_A} \,,
\\ 
\label{mu}
& 
& 
\hspace{-.7cm}
\mbox{the energy-to-mass flux ratio} 
\
\mu(A) c^2=\xi \gamma c^2 - \frac{\varpi \Omega B_\phi}{\Psi_A }\,,
\\ 
\label{adiabat}
& 
& 
\hspace{-.7cm}
\mbox{the adiabat}
\
Q(A)=\frac{P}{\rho_0^\Gamma} \,,
\end{eqnarray}
\end{mathletters}
\noindent
where subscripts $p$/$\phi$ denote poloidal/azimuthal components.

An important combination of the above field line constants is the ``Michel's 
magnetization parameter''
\begin{equation}\label{sigmaM}
\sigma_{\rm M}(A) \equiv
\frac{A \Omega^2}{\Psi_A c^3}\,.
\end{equation}

The physical quantities can be expressed as
functions of $A$ and the ``Alfv\'enic'' Mach number
$M\equiv ( \gamma V_p / B_p)(4 \pi \rho_0 \xi)^{1/2} $:
\begin{mathletters}\label{quantities}
\begin{equation}\label{rho}
\rho_0=\frac{\xi \Psi_A^2}{4 \pi M^2} \,,
\end{equation}
\begin{equation}\label{gamma}
\gamma=\frac{\mu}{\xi} \frac{M^2-(1-x_{\rm A}^2)}{M^2+x^2-1} \,,
\end{equation}
\begin{equation}\label{B-E}
{\boldsymbol{B}}=\frac{{\boldsymbol{\nabla}}A \times \hat{\phi} }{\varpi}
-\frac{c \Psi_A}{x} \left(\mu - \xi \gamma \right) \hat{\phi} 
\,, \quad
{\boldsymbol{E}}=-\frac{\Omega}{c} {\boldsymbol{\nabla}}A\,,
\end{equation}
\begin{equation}
\frac{V_p}{c} = \frac{M^2 B_p}{\xi \gamma \Psi_A c}
\,, \quad
\frac{V_\phi}{c} = 
\frac{\xi \gamma - \mu \left(1- x_{\rm A}^2 \right) }{\xi \gamma x} 
\label{V}
\end{equation}
\end{mathletters}
\noindent
Here $x=\varpi \Omega / c$ is the cylindrical distance in units
of the light cylinder's lever arm on each field-streamline $A=$const,
and $x_{\rm A}=\left({L \Omega}/{\mu c^2}\right)^{1/2}$ 
is its value at the Alfv\'en point.\footnote{
We use the term ``light cylinder'' for the surface $\varpi \Omega (A) =c$, although
in the general case where $d\Omega / dA \neq 0 $ it is not a cylinder.} 

Alternatively, using the equivalent to equation (\ref{gamma}),
\begin{equation}\label{m2}
M^2=\frac{ \xi \gamma }{\mu - \xi \gamma} x^2 -
\frac{ \xi \gamma - \mu \left(1- x_{\rm A}^2 \right) }{\mu - \xi \gamma} 
\end{equation}
and the Poynting-to-matter energy flux ratio
\begin{equation}\label{sigma}
\sigma \equiv - \frac{\varpi \Omega B_\phi}{\Psi_A \xi \gamma c^2} 
=\frac{\mu - \xi \gamma}{ \xi \gamma} 
\ \Leftrightarrow \ \xi \gamma = \frac{\mu}{1+\sigma}
\,,
\end{equation}
we may rewrite all quantities as functions of ($A\,, \gamma$) or ($A\,, \sigma$).

The two remaining equations are the Bernoulli
and the transfield force-balance relations.
The identity
$\gamma^2 - 1 = \left(\gamma V_p/c \right)^2 +\left(\gamma V_{\phi}/c \right)^2$
gives the Bernoulli equation
\begin{eqnarray}\label{bernoulli}
&&\gamma^2 - 1 =
\nonumber \\ && =
\Sigma^2 \left[
\frac{ \mu \gamma }{\mu - \xi \gamma} -
\mu \frac{ \xi \gamma - \mu \left(1- x_{\rm A}^2 \right) }{\xi \left(\mu - \xi \gamma \right) x^2}
\right]^2
+\left[
\frac{\xi \gamma - \mu \left(1- x_{\rm A}^2 \right) }{ \xi x}
\right]^2
\end{eqnarray}
where
\begin{equation}\label{Sigma}
\Sigma \equiv 
\frac{\Omega^2 \varpi | {\boldsymbol{\nabla}} A | }
{\mu \Psi_A c^3 }
\quad \left( = \frac{\sigma_{\rm M}}{\mu} \frac{ \varpi^2 B_p}{A} \right)
\,.
\end{equation}
The projection of equation (\ref{momentum})
along $\hat{n}\equiv - {\boldsymbol{\nabla}} A / | {\boldsymbol{\nabla}} A |$
gives the transfield force-balance equation
\begin{eqnarray}
{f}_{C \bot} + {f}_{I \bot} + {f}_{P \bot} 
+{f}_{EM1} +{f}_{EM2}+{f}_{EM3}=0 \,,
\nonumber
\end{eqnarray}
\noindent
where:
\begin{itemize}
\item
[--]
The azimuthal centrifugal term
\begin{eqnarray}
{f}_{C \bot}=
\xi \gamma^2 \rho_0 \frac{V_\phi^2}{\varpi} \hat{n} \cdot \hat{\varpi} =
\frac{B_p^2}{4 \pi \varpi} \left(\frac{M V_{\phi} }{V_p}\right)^2 \hat{n} \cdot
 \hat{\varpi} \,.
\nonumber
\end{eqnarray}
\item
[--]
The rest of the inertial force density along $\hat{n}$
\begin{eqnarray}
{f}_{I \bot}=
-\gamma^2 \rho_0 \xi \hat{n} \cdot
\left[\left({\boldsymbol{V}} \cdot {\boldsymbol{\nabla}} \right) {\boldsymbol{V}} \right]
-{f}_{C \bot}
=-\frac{B_p^2}{4 \pi {\cal R}} M^2
\nonumber
\end{eqnarray}
is the poloidal centrifugal term. Here
${\cal R}$ is the curvature radius of the poloidal field lines.
\item
[--]
The pressure gradient force density along $\hat{n}$
\begin{eqnarray}
{f}_{P \bot}=-\hat{n} \cdot {\boldsymbol{\nabla}} P
\,.
\nonumber
\end{eqnarray}
\item
[--]
The ``electric field'' force density
\begin{eqnarray}
{f}_{E \bot}=\frac{1}{ 8 \pi \varpi^2 } \hat{n} 
\cdot {\boldsymbol{\nabla}} \left( \varpi^2 E^2 \right)
-\frac{{E^2}}{{4 \pi {\cal R}} } 
\,.
\nonumber 
\end{eqnarray}
\item
[--]
The ``magnetic field'' force density  along $\hat{n}$
\begin{eqnarray}
{f}_{B \bot}=-\frac {1}{8 \pi \varpi^2 } 
\hat{n} \cdot {\boldsymbol{\nabla}} \left( \varpi^2 B^2 \right)
+ \frac{{B_p^2}}{{4 \pi {\cal R}} } 
+\frac{B_p^2}{4 \pi \varpi} \hat{n} \cdot \hat{\varpi}\,.
\nonumber
\end{eqnarray}
\end{itemize}
The total electromagnetic force density in the transfield direction 
$({f}_{E \bot}+{f}_{B \bot})$ can be decomposed as
\begin{equation}
\underbrace{
-\frac {1}{8 \pi \varpi^2 } \hat{n} 
\cdot {\boldsymbol{\nabla}} \left[ \varpi^2 \left(B^2 -E^2 \right) \right] }_{f_{EM1}}
+\underbrace{
\frac{{B_p^2(1-x^2)}}{{4 \pi {\cal R}} } }_{f_{EM2}}
+\underbrace{
\frac{B_p^2}{4 \pi \varpi} \hat{n} \cdot \hat{\varpi}}_{f_{EM3}}
\end{equation}
Altogether, they give the following form of the transfield force-balance equation:
\begin{eqnarray}\label{transfield}
\frac{B_p^2}{4 \pi {\cal R}}\left(M^2+x^2-1\right)=
-\frac{1}{8 \pi \varpi^2} \hat{n} \cdot {\boldsymbol{\nabla}} 
\left[\varpi^2 \left(B^2 -E^2 \right) \right]
+ \nonumber \\
+\frac{B_p^2}{4 \pi \varpi}
\left[1+ \left(\frac{M V_{\phi} }{V_p}\right)^2 \right] \hat{n} \cdot \hat{\varpi}
- \hat{n} \cdot {\boldsymbol{\nabla}} P
 \,.
\end{eqnarray}
(Using equations [\ref{quantities}], the latter becomes a second-order 
partial differential equation (PDE) for $A$.)

At the origin of the flow (subscript $i$),
$x_i \ll x_{\rm A}$, $M_i^2 \ll 1-x_{\rm A}^2$,
and (using eq. [\ref{gamma}])
$\gamma_i \xi_i (1-x_i^2) \approx \mu (1-x_{\rm A}^2)$.
As the flow moves downstream, the functions $x$ and $M$ increase,
and when the Alfv\'en surface is reached they become $x=x_{\rm A}$, $M^2=1-x_{\rm A}^2$.
The pulsar wind is still strongly magnetized
in the neighborhood of the Alfv\'en surface, 
and the field is close to force-free there:
$| {f}_{I \bot} | \ll | {f}_{B_p \bot}|$,
or, $M^2 \ll 1$, implying $x_{\rm A} \lesssim 1$.\footnote{
This is the case for trans-Alfv\'enic flows only. If the flow starts
with super-Alfv\'enic velocity, $x_{\rm A}$ could be in general larger than unity.}
The light cylinder $x=1$ is located slightly after the Alfv\'en surface.
Next, the flow enters the super-Alfv\'enic regime.

\subsection{The super-Alfv\'enic asymptotic regime}
\label{superalfvenic}
In this regime $x^2\gg x_{\rm A}^2 \approx 1$, $M^2 \gg 1-x_{\rm A}^2$.
Also thermal effects play no longer any role and the fluid is cold, $\xi \approx 1$.
It is reasonable to assume that the above inequalities hold from the 
classical fast-magnetosound point downstream.
At this point $(\gamma V_p)^2 \approx (B^2-E^2)/4 \pi \rho_0$,
and for cold flows,  $\gamma \approx \mu^{1/3} \gg 1$
(e.g., \citealt{C86}).\footnote{
Note that the flow is cold at the classical fast-magnetosound
point when $\gamma_i \xi_i \ll \mu^{1/3}$, or, 
(using eq. [\ref{gamma}] with $x_i\ll x_{\rm A}$, $M_i^2 \ll 1-x_{\rm A}^2$),
$1-x_{\rm A}^2 \ll (1-x_i^2) \mu^{-2/3}$, implying $x_{\rm A} \lesssim 1$.}
\\
Employing the comoving magnetic field $B^2-E^2=B_p^2(1-x^2) + B_\phi^2$,
the Bernoulli equation (\ref{bernoulli}) gives the exact result
\begin{equation}\label{bernoulli-comB}
B^2-E^2= \frac{x^2\left[x^2-1+\mu^2 
(1-x_{\rm A}^2)^2/\xi^2\right]}{\left[x^2-1+\mu (1-x_{\rm A}^2) / \xi \gamma \right]^2}
\frac{B_\phi^2}{\gamma^2} \,.
\end{equation}
In the asymptotic regime where $x\gg 1$ the latter equation implies
\begin{equation}\label{bpbphi}
B^2-E^2 \approx \frac{B_\phi^2}{\gamma^2} \,,  \ 
\frac{x^2 B_p^2}{ B_\phi^2} 
=\frac{B_\phi^2-\left(B^2-E^2\right)}{B_\phi^2 \left(1-1/x^2 \right)} 
\approx \frac{1-1/\gamma^2}{1-1/x^2}\,.
\end{equation}
Thus, the magnetic field is mainly azimuthal and the electric field
approximately equals the magnetic field\footnote{
In cases where $E$ {\emph {exactly}} equals $-B_\phi$,  eq. (\ref{Eapprox})
implies a ``linear accelerator'' $\gamma = x$ \citep{CK02}.
However, in the general case $E$ does not {\emph {exactly}} equals $-B_\phi$,
and eq. (\ref{Eapprox}) gives just the small difference between them.
Note that the $E=-B_\phi$ is not exactly satisfied in the numerical solution
of \citet*{CKF99} either.}
\begin{equation}\label{Eapprox}
E = \frac{\Omega}{c} | {\boldsymbol{\nabla}} A | =
 xB_p \approx -B_\phi \left(
 \frac{1-1/\gamma^2}{1-1/x^2}
 \right)^{1/2} \,.
\end{equation}
The approximate form of the Bernoulli equation (\ref{bernoulli}) is
\begin{equation}\label{bernoulliapprox}
\left({\gamma^2-1}\right)^{1/2} \approx 
\Sigma
\frac{ \mu \gamma }{\mu - \gamma} \,.
\end{equation}

We are now in a position to examine the highly nonlinear transfield 
force-balance equation (\ref{transfield}).

In that equation, the ${f}_{B \bot}$ and ${f}_{E \bot}$ are the dominant terms 
that almost cancel each other (e.g. \citealt{B01,VK03a}).
The algebraic sum of their main contributions is the $f_{EM1}$ term, which
can be written as (using eq. [\ref{bpbphi}])
$-(1/8 \pi \varpi^2) \hat{n} \cdot {\boldsymbol{\nabla}} \left( 
\varpi B_{\phi} / \gamma \right)^2$.
This term must be balanced by the other terms of equation (\ref{transfield}).
Ignoring ${f}_{P \bot} $, we are left with the ``poloidal curvature'' term
$(B_p^2 / 4 \pi {\cal R} ) (M^2+x^2-1) = -f_{I \bot}-f_{EM2}$
and the term
$-({B_p^2}/{4 \pi \varpi}) [1+ ({M V_{\phi}}/{V_p})^2] \hat{n} \cdot \hat{\varpi}
=-f_{C \bot} -f_{EM3}$. Note that the last term, which henceforth we call ``centrifugal'',
has contributions from the poloidal magnetic field (the part $-f_{EM3}$ which is the dominant
one in the $\sigma \gg 1$ regime), and from the azimuthal centrifugal force $-f_{C \bot}$.
(The ratio of the two ``centrifugal'' parts is
$f_{EM3}/ f_{C \bot} \approx M^2/x^2 \approx \sigma$,
and the term $f_{C \bot}$ becomes important in the matter-dominated regime.)
Thus, the asymptotic form of the transfield force-balance equation (\ref{transfield}) is
\begin{equation}\label{transfieldapprox}
(M^2+x^2) \frac{\varpi}{ {\cal R} }  \approx
-\frac{\hat{n} \cdot {\boldsymbol{\nabla}} }{2 \varpi B_p^2 }
\left[ \frac{\varpi B_{\phi} }{\gamma} \right]^2
+\left[ 1 + \frac{M^2 \gamma^2 }{x^2 (\gamma^2-1)} \right] 
\hat{n} \cdot \hat{\varpi} \,.
\end{equation}
This important equation, with the ``centrifugal'' term omitted, was derived by 
\citet{CLB91,LE01,O02}, while recently \cite{TT03} derived the same equation with 
the ``centrifugal'' term included, but the poloidal curvature term omitted.

It is convenient to rewrite the Bernoulli and transfield force-balance equations
in terms of the unknown functions ($A\,, \sigma$)
(also assuming ${\gamma^2-1}\approx \gamma^2$).
The Bernoulli equation (\ref{bernoulliapprox}) becomes\footnote{
To a better approximation equation (\ref{bernoulliapprox}) yields
$\mu \Sigma \approx \mu -\gamma - \mu / 2 \gamma^2$.
The last term on the right-hand side is nonnegligible near the classical fast-magnetosound
surface in the sense that the derivative of the latter
equation (including the last term) along the flow gives
$\left(\gamma^{-3}-\mu^{-1}\right) {\boldsymbol V} \cdot {\boldsymbol {\nabla}} \gamma=
{\boldsymbol V} \cdot {\boldsymbol {\nabla}} \Sigma$, implying $\gamma_{f}=\mu^{1/3}$,
$\Sigma_{f}=1-3/2 \mu^{2/3}$.
However, $\mu / 2 \gamma^2 \ll \mu$ and we neglect this term in our analysis.}
$\gamma = \mu \left(1- \Sigma \right)$, 
or (using eqs. [\ref{sigma}], and [\ref{Sigma}]),
\begin{mathletters}\label{systemfinal}
\begin{equation}\label{bernoullifinal}
\frac{\sigma}{1+\sigma} = 
\frac{\sigma_{\rm M}}{\mu} \frac{ \varpi | {\boldsymbol{\nabla}} A |}{A} 
\quad \left( = \frac{\sigma_{\rm M}}{\mu} \frac{ \varpi^2 B_p}{A} \right)
\,.
\end{equation}
The transfield force-balance equation (\ref{transfieldapprox}) becomes
(using eqs. [\ref{B-E}], [\ref{m2}], [\ref{sigma}], and [\ref{bernoullifinal}])
\begin{equation}\label{transfieldfinal}
\frac{\varpi}{{\cal R}}=
\frac{ \sigma (1+\sigma) \varpi
{\boldsymbol {\nabla}} A
\cdot {\boldsymbol {\nabla}} 
} { \mu^2 | {\boldsymbol {\nabla}} A |} 
\ln \frac{A \Omega \sigma }{\sigma_{\rm M} } -
\frac{c^2 }{\varpi^2 \Omega ^2 }
\frac{\hat{\varpi} \cdot {\boldsymbol {\nabla}} A} { | {\boldsymbol {\nabla}} A |}
\,.
\end{equation}
Equations (\ref{bernoullifinal}), and (\ref{transfieldfinal}) together with the equation
for the curvature radius,
\begin{equation}
{\cal R}= \frac
{| {\boldsymbol{\nabla}} A | } 
{{\boldsymbol{\nabla}}^2 A
- {\boldsymbol{\nabla}} A \cdot {\boldsymbol{\nabla}}
\ln | \varpi {\boldsymbol{\nabla}} A | }
\,,
\end{equation}
form a closed system for the functions $\sigma$ and $A$.
A possible solution of the system will describe the outflow 
downstream from the classical fast-magnetosound surface
(where the super-Alfv\'enic asymptotic conditions surely hold).
\end{mathletters}

Our task is to find a field geometry $A(\varpi\,, z)$, such that the flow
accelerates smoothly from $\sigma \gg 1$ (at the classical fast-magnetosound surface)
to $\sigma \ll 1$ (in the matter-dominated regime).
Equation (\ref{bernoullifinal}) shows that the key quantity
$\varpi | {\boldsymbol{\nabla}} A | = \varpi^2 B_p$ 
should be a decreasing function of the distance from the source,
along any field line.
According to the model of Kennel \& Coroniti (1984a,b) for the Crab nebula,
the solution should give $\sigma_{s} \approx 0.003$, 
$\gamma_{s} \approx \mu \approx 10^6$, and $B_{\phi,{s}} \approx -2 \times 10^{-5}$G,
at the distance $r_{s} \approx 3 \times 10^{17}$ cm of the termination shock.
At the classical fast-magnetosound surface 
$\gamma_{f} \approx \mu^{1/3} = 10^2$ and, from equation (\ref{sigma}),
$\sigma_{f} \approx 10^4$.

\subsubsection{General solution characteristics}\label{character}

\qquad \underline{
The current distribution}\\
The number density of the particles in the pulsar magnetosphere 
highly exceeds the Goldreich-Julian value,
and a tiny deviation from neutrality is enough to support the electromagnetic field.
Assuming that in the upper (lower) hemisphere $B_\phi < 0 (>0)$,
the Poynting-flux is outflowing when
the electric field points toward the rotation axis, corresponding to a negative value
of the charge density in the polar regions ($J^0<0$).
For a highly relativistic poloidal motion,
$$J^0=\frac{c}{4 \pi} {\boldsymbol{\nabla}} \cdot {\boldsymbol{E}}=
-\frac{1}{4 \pi} {\boldsymbol \nabla}
\left({\boldsymbol{V}} \times {\boldsymbol{B}} \right) \approx
\frac{1}{4 \pi} {\boldsymbol{V}}_p \cdot 
{\boldsymbol{\nabla}} \times {\boldsymbol{B}} \approx J_\parallel $$ 
(where a suffix ``$\parallel$'' denotes the component of a vector
along the poloidal field),
and the flow that originates from the negatively-charged polar region is 
current-carrying ($J_\parallel <0$).\footnote{For
the assumed quasi-neutral plasma, the sign of $J_\parallel$ depends on the tiny relative
speed between the positively and negatively charged species.}
The current-closure condition demands that
each field line reaches a point where $J_\parallel=0$, and after that enters the
distributed ``return-current'' regime $J_\parallel > 0$.
The poloidal current density is ${\boldsymbol {J}}_p=\frac{c}{4 \pi } {\boldsymbol {\nabla}}
\times {\boldsymbol {B}}_{\phi} = \frac{1}{2 \pi \varpi}
{\boldsymbol {\nabla}} I \times \hat{\phi}$,
with $I=\int \! \! \! \! \int {\boldsymbol{J}}_p \cdot d {\boldsymbol {S}}
=\frac{c}{2} \varpi B_{\phi}$, and the meridional current lines
represent the loci of constant total poloidal current ($I=const$). 
As a flow element crosses successive current lines,
the absolute value of $I$ decreases and asymptotically vanishes ($I_\infty=0$).
\\
An alternative scenario is that the current closes in a thin sheet
at the end of the ideal MHD asymptotic regime, in which case
$I_\infty = I_\infty (A) \neq 0$.

\qquad \underline{
The $\sigma \sim 1$ transition}\\
The field is force-free near the pulsar and the
energy is Poynting flux-dominated: $\sigma \gg 1$.
This remains the case near the classical fast-magnetosound surface, where 
$\sigma \approx \mu /\gamma_{f} \approx \mu^{2/3} \gg 1$.
As the Lorentz force accelerates the matter -- meaning that the flow crosses
current lines with decreasing $| I |$ -- the function $\sigma$ also decreases.
The relation between $\sigma$ and $I$ is (by combining equations
[\ref{mu}], [\ref{sigma}])
\begin{equation}\label{currdistr}
\frac{\sigma}{1+\sigma}=\frac{2(-I)}{A \Omega}  \frac{\sigma_{\rm M}}{\mu}\,.
\end{equation}
At $\sigma \sim $ a few the force-free approximation brakes down and
the back-reaction of inertial forces on the magnetic field becomes important.
The final value of $\sigma$ could be in general very small (or exactly zero),
depending on the small (or zero) value of $I_\infty$.

\qquad \underline{
The $\sigma >\sigma_{c} $ regime where ${f}_{C \bot} + f_{EM3}$ is negligible}\\
The ``centrifugal'' term in the transfield equation (the last term in eq. [\ref{transfieldfinal}])
is negligible in the force-free regime.
Since the flow velocity is mainly poloidal, it is expected that this term is also negligible
inside the matter-dominated regime,
in which case the poloidal curvature is controlled by the electromagnetic force $f_{EM1}$.
Nevertheless, as the flow becomes more and more matter-dominated, the electromagnetic
force $f_{EM1}$ decreases (it vanishes at $\sigma=0$).
As a result, a point $\sigma = \sigma_{c}$ is reached where 
the two terms on the right-hand side of 
equation (\ref{transfieldfinal}) become comparable,
and after that point the azimuthal centrifugal term takes over. 
\\
For $\sigma > \sigma_{c}$ -- when the centrifugal term is negligible --
equation (\ref{transfieldfinal}) yields (using eqs. [\ref{sigmaM}], [\ref{bernoullifinal}])
\begin{equation}\label{sintheta}
\left(\frac{d \cos \vartheta}{d \ln \varpi}\right)_A=
\frac{ \sigma c^3} {\mu \Omega}
\frac{{\boldsymbol {\nabla}} A
\cdot {\boldsymbol {\nabla}}
} { | {\boldsymbol {\nabla}} A |^2} 
\frac{\sigma \Psi_A}{\Omega }
\,.
\end{equation}
Here $\vartheta$ is the opening half-angle of the flow
(the angle between the poloidal magnetic field and the rotation axis).
\\
If the flow reaches a constant value 
(in principle different from field line to field line) $\sigma_\infty(A)$ such that
$\sigma_\infty > \sigma_{c}$, then the right-hand side of equation 
(\ref{sintheta}) is a function
of $A$ alone and $\cos \vartheta$ would diverge logarithmically with $\varpi$ \citep{O02}.
The only way to avoid this divergence is to have
$\sigma_\infty(A) \Psi_A (A)/ \Omega(A)=$const.
By using equations (\ref{B-E}), and (\ref{sigma})
the solvability condition at infinity $\gamma_\infty(A) / I_\infty(A)
=$const (Heyvaerts \& Norman 1989; Chiueh et al. 1991) is recovered.

\qquad \underline{
The $\sigma <\sigma_{c} $ regime where ${f}_{C \bot}$ is important}\\
At sufficiently small values of $\sigma$ ($ <\sigma_{c} $),
the curvature of the poloidal field-streamlines is controlled by 
the azimuthal centrifugal force\footnote{
The other part of the ``centrifugal'' force, namely the $f_{EM3}$, equals 
$\sigma {f}_{C \bot} \ll {f}_{C \bot}$
and can be omitted in the matter-dominated regime.};
the electromagnetic force is much smaller than its inertial counterparts,
and equation (\ref{transfieldfinal}) gives 
\begin{equation}\label{hyperregime}
\frac{\varpi}{ {\cal R}} = -\frac{\cos \vartheta}{ x^2} \quad (< 0) \,.
\end{equation}
In other words, in the $\sigma <\sigma_{c} $ regime we cannot neglect
terms of order ${\cal O} \left(x_{\rm A}^2/x^2\right)$, ${\cal O} \left(\varpi/ {\cal R}\right)$ in the
transfield equation, simply because the electromagnetic term is even smaller.
\\
In this regime the solution of the momentum equation is simply
${\boldsymbol {V}}\approx$const and the motion is ballistic
(in the poloidal plane the azimuthal and poloidal centrifugal forces
cancel each other and the total inertial force vanishes).
By decomposing a straight in three dimensions, constant velocity streamline
in cylindrical coordinates,
we get: $V_z=const$, $V_\phi=c/x=c^2 / \varpi \Omega$ (from angular momentum conservation),
and an increasing $V_\varpi=\left(V^2-V_z^2-c^2/x^2\right)^{1/2}$.
It can be shown that the poloidal streamline shape 
(the projection of the straight in three dimensions streamline onto the poloidal plane) 
is hyperbolic,
$$ \frac{V^2-V_z^2}{c^2}\left(\frac{\varpi \Omega }{c}\right)^2-\frac{(V^2-V_z^2)^2}{V_z^2 c^2}
\left(\frac{(z-z_0) \Omega }{c} \right)^2 =1 \,,$$
with some constant $z_0$.
\\
We emphasize that for flows with $\sigma_\infty=\sigma_\infty(A)<\sigma_{c}$, 
equation (\ref{sintheta})
should be replaced by equation (\ref{hyperregime})
and there is no problem related to the divergence of $\cos \vartheta$.
In the absence of a solvability condition at infinity,
any value $\sigma_\infty(A)<\sigma_{c}$ 
(and the corresponding current distribution, see eq. [\ref{currdistr}])
is a possible solution.
Even so, the most plausible case is
$\sigma_\infty=0$ corresponding to a distributed return-current regime
with $I_\infty=0$ \citep{O02}.

\qquad \underline{
The modified fast-magnetosound singular surface}\\
The classical fast-magnetosound surface is not singular when 
we solve simultaneously the Bernoulli {\emph{and}} the transfield force-balance equations
(it just separates the elliptic and hyperbolic regimes of the PDE problem).
The singular surface is the so-called modified fast-magnetosound surface
(or fast-magnetosound separatrix surface) that lies inside the hyperbolic regime,
coincides with a limiting characteristic, and plays the role of the event horizon 
for the propagation of fast waves \citep{TSSTC96,B97,VK03a}.
The causality principle is satisfied (disturbances in the asymptotic regime cannot influence
the flow near the origin) only if the solution is trans-modified fast.
Completely matter-dominated flows will have super--modified fast-magnetosound velocities
(simply because the fast-magnetosound speed vanishes as $\sigma \rightarrow 0$).

\section{The $z$ self-similar model}\label{zss}

Assuming that the poloidal field line shape is approximately cylindrical, i.e.,
$B_\varpi \ll B_z \Leftrightarrow
| \partial A / \partial z | \ll \partial A / \partial \varpi$,
we may further simplify the system of equations (\ref{systemfinal}) to
\begin{mathletters}\label{system_cyl}
\begin{eqnarray}
\frac{\sigma}{1+\sigma}
&=& \frac{\sigma_{\rm M}}{\mu} \frac{\varpi}{A}  \frac{\partial A}{\partial \varpi}
\,, \\
\frac{\varpi}{{\cal R}} &=&
\frac{\sigma (1+\sigma) \varpi }{\mu^2 }
\frac{\partial }{\partial \varpi}
\ln \left( \frac{A \Omega \sigma }{\sigma_{\rm M} } \right)
-\frac{c^2 }{ \varpi^2 \Omega^2 }
\,.
\end{eqnarray}
\end{mathletters}

\begin{mathletters}
To construct classes of analytical solutions, we shall make two assumptions:
$z$ is assumed a product of a function of $A$ times a function of $\varpi$ 
(the $z$ self-similar ansatz)
\begin{equation}\label{z_cyl}
z = \varpi_0 \Phi (A) \ \zeta(\varpi)
\end{equation}
(where $\varpi_0=$const), and 
\begin{equation}
\sigma = \sigma (\varpi)\,.
\end{equation}
\end{mathletters}

Following the algorithm described in \citet{VT98},
it is possible to separate the variables $A$ and $\varpi$ in the system (\ref{system_cyl})
only if the following relations hold:
\begin{mathletters}\label{flux1}
\begin{eqnarray}
A-A_0&=&\frac{ \tau}{(-F)} B_0 \varpi_0^2 \Phi^F\,,
\label{flux}
\\
\mu &=& \mu_0 \Phi \,,
\\
\Omega &= & \frac{c}{\varpi_0} \Phi\,,
\label{varpi_0}
\\
\frac{\mu}{\sigma_{\rm M}}&=& \frac{(-F)}{\tau} \frac{A-A_0}{A}\,.
\end{eqnarray}
Here $F\,, B_0\,, \mu_0\,,$ and $\tau$ are constants.
\end{mathletters}
The system (\ref{system_cyl}) becomes
\begin{mathletters}
\begin{eqnarray}
 \frac{\sigma}{1+\sigma}
&=& \tau \varpi \frac{d \zeta }{ \zeta d \varpi} \,, \\
\frac{\varpi}{{\cal R}} &=&
 \frac{\varpi_0^2}{z^2} \frac{(1+\sigma)  \zeta^{F+2} \varpi }{\mu_0^2 }
\frac{d \left( \sigma  \zeta^{-F} \right) }{d \varpi } 
-\frac{\varpi_0^4  \zeta^2 }{ \varpi^2 z^2 }
\,.
\label{trans1}
\end{eqnarray}
\end{mathletters}
Using the expression for the curvature radius
\begin{equation}\label{curvature_cyl}
\frac{\varpi}{{\cal R}} = 
\frac{\varpi }{z^2}  \zeta^2 \frac{d^2  \zeta}{d \varpi^2} 
\left(\frac{d  \zeta}{d \varpi}\right)^{-3}
\left(1+\frac{B_\varpi^2}{B_z^2}\right)^{-3/2} \,,
\end{equation}
we get finally a system of ordinary differential equations (ODEs)
\begin{mathletters}\label{ode_cyl}
\begin{eqnarray}
\label{ode_cyl1}
&& 
\frac{d \varpi}{d  \zeta } = \frac{\tau \varpi (1+\sigma) }{ \zeta \sigma}  \,, \\ &&
\frac{d \sigma}{d  \zeta}=
\left[\frac{\mu_0^2 \tau^2 \varpi^2 }{\varpi_0^2  \zeta^3 }                                     
\frac{1+\sigma }{\sigma } \left(\tau \frac{1+\sigma }{\sigma }  - 1 \right)
-\frac{F \sigma^2}{ \zeta} - \frac{\mu_0^2 \varpi_0^2 \tau }{  \zeta \varpi^2 } \right]
\nonumber \\ &&
\quad \ \ \times 
\left[ \frac{\mu_0^2 \tau^2 \varpi^2 } {\varpi_0^2  \zeta^2 \sigma^2 }-            
\sigma \right]^{-1} \,.
\label{ode_cyl2}
\end{eqnarray}
\end{mathletters}
The latter system can be easily integrated; the only difficulty is that the solution
should cross a singular point that appears when the denominator in
equation (\ref{ode_cyl2}) vanishes.
By rewriting this denominator as
\begin{equation}
\frac{\mu_0^2 \tau^2 \varpi^2 } {\varpi_0^2  \zeta^2 \sigma^2 }- \sigma =
\left(\frac{\gamma V_\varpi}{c}\right)^2 - \frac{B^2-E^2}{4 \pi \rho_0 c^2 } \,,
\end{equation}
it is evident that the singular point corresponds to the
modified fast-magnetosound point, where the phase speed of the
fast-magnetosound waves propagating along $\hat \varpi$ is zero 
(see Appendix C in \citealt{VK03a}).

Using equation (\ref{z_cyl}) we may find ${\boldsymbol \nabla} A$
and hence the magnetic field,
\begin{equation}\label{bifield}
{\boldsymbol {B}} = \frac{B_0 \varpi_0^2 \Phi^{F} }{ \varpi^2 }
\left( \frac{\sigma}{1+\sigma} \hat{z} + \tau \frac{\varpi}{z} \hat{\varpi} \right)
-\frac{B_0 \varpi_0 \Phi^{F+1} }{ \varpi }\frac{\sigma}{1+\sigma} \hat{\phi} \,,
\end{equation}
whereas the current density component $J_\parallel$ is given by
\begin{equation}\label{jparallel}
J_\parallel \approx J_z = \frac{c B_\phi}{4 \pi}
\frac{\sigma}{1+\sigma} \left( -\frac{F+1}{\tau}-
\frac{\varpi d \sigma^{-1}}{d \varpi}\right) \,.
\end{equation}
The magnetic flux function is 
\begin{equation}
\label{flux2}
 A - A_0 =
\frac{\tau (1+\sigma) }{(-F) \sigma} B_p \varpi^2
\,. 
\end{equation}

\subsection{Results}
\subsubsection{Scaling laws}\label{scalings}
It is instructive to derive simple analytical scalings from equations (\ref{ode_cyl}):
they turn out to be in very good agreement with the results of the numerical integration
presented below.
Before analyzing separately each regime, we rewrite schematically equation (\ref{ode_cyl2}) as
\begin{equation}\label{schemasigma}
\frac{d\sigma}{ d \zeta} = \frac{ -{\cal {R}}_1 + 
{\cal {B}}_1 - {\cal {C}}}{{\cal {R}}_2-{\cal {B}}_2 } \,.
\end{equation}
All the terms appearing on the right-hand side are positive (for $\sigma > \tau$ and $F<0$).
It is important to note that the terms ${\cal {R}}_1$, ${\cal {R}}_2$ come from the
poloidal curvature term of the transfield equation, the ${\cal {B}}_1$ and ${\cal {B}}_2$ have 
electromagnetic origin, while ${\cal {C}}$ comes from the ``centrifugal'' term.

\qquad \underline{
The sub-modified fast regime $\sigma \gg \sigma_{mf}$}\\
In this regime the denominator in equation (\ref{schemasigma})
is negative, ${\cal {R}}_2 < {\cal {B}}_2 $, i.e., the term ${\cal {B}}_2$ dominates.
Since $d\sigma /d \zeta < 0$, the numerator should be positive. Thus, 
the dominant term is the ${\cal {B}}_1$ (the only positive term).
As we expected, in the sub-modified fast regime the magnetic field dominates
over the inertial terms. One may write $d\sigma /d \zeta \approx -{\cal {B}}_1/{\cal {B}}_2 $.
The integration gives $ \sigma \propto \zeta^F$ and we finally have (by employing eq. [\ref{ode_cyl1}])
\begin{equation}\label{regime1}
z \propto \Phi \varpi^{1/\tau}\,, 
\quad
\sigma \propto \Phi^F z^F\,.
\end{equation}
Equation (\ref{jparallel}) gives
$J_\parallel \approx - \frac{c B_\phi}{4 \pi} \frac{F+1}{\tau}$.
Thus, for $F<-1$ the flow is current-carrying ($J_\parallel<0$) near the origin.

\begin{figure*}[t]
  {\includegraphics[scale=.93]{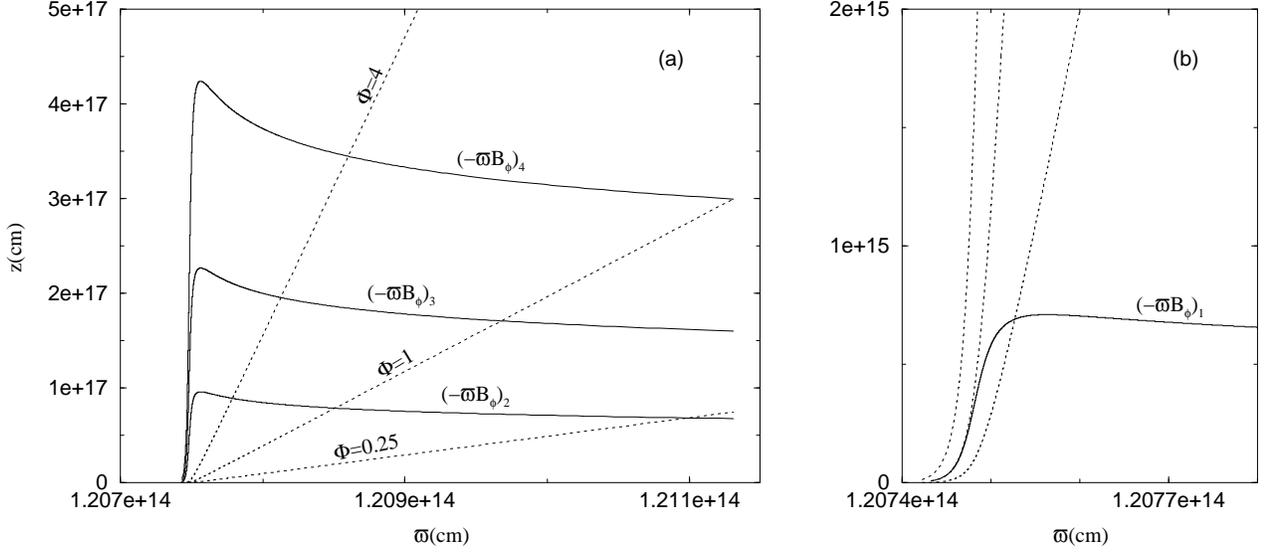}}
  \caption{
Solution $a$, corresponding to 
$\tau=10^{-5}$ and $F=-1.9$.
($a$) Three field lines ($\Phi=0.25\,, 1\,, 4$ -- {\emph {dashed}} lines) 
and four poloidal current lines 
[$(-\varpi B_\phi)_{1\,,2\,,3\,,4}=
9 \times 10^{11}\,,10^{10}\,, 5 \times 10^{9}\,, 3 \times 10^{9}$cgs --
{\emph {solid}} lines] for the solution $a$.
As the flow moves along a particular field line it crosses current lines with 
decreasing $| I | =(c/2) (-\varpi B_\phi )$, 
meaning that it is accelerated and the Poynting-to-matter energy flux ratio decreases.
The field line shape is close to conical; see eq. (\ref{regime2}). 
The current-carrying regime $J_\parallel <0$ near the rotation axis, and 
the return-current regime $J_\parallel >0$ at larger cylindrical distances are shown.
($b$) The same field lines with an enlarged scale,
focusing on the region close to the classical fast-magnetosound surface.
The field lines have ${\cal R}>0$ and the shape is parabolic; see eq. (\ref{regime1}).
A characteristic of the force-free regime is ${\boldsymbol{B}}_p \parallel {\boldsymbol{J}}_p$,
e.g., the current line $-\varpi B_\phi=(-\varpi B_\phi)_1$
is initially tangent to the middle field line $\Phi=1$.
(This is the case for $z<3\times 10^{14}$cm, for which fig. \ref{fig2} shows that $\sigma > 9$.)
\label{fig1}}
\end{figure*}

\qquad \underline{
The super-modified fast regime $\sigma \ll \sigma_{mf}$}\\
In this regime, ${\cal {R}}_2 > {\cal {B}}_2 $. Since $d\sigma /d \zeta < 0$,
the inertial terms dominate in the numerator as well,
resulting in $d\sigma /d \zeta \approx -({\cal {R}}_1+{\cal {C}})/ {\cal {R}}_2$.
As long as $\sigma > \sigma_{c}$ we get that
not only the poloidal curvature terms are important and should not be neglected, but
also that the transfield force-balance equation reduces to $\varpi / {\cal R}=0$.
In other words,
it is not correct to substitute $\varpi / {\cal R} \approx 0 $ 
(the ``pseudo-force-free'' condition according to \citealp{O02})
in the transfield equation and keep only the other terms. 
On the contrary, {\emph{the $\varpi / {\cal R} \approx 0 $
is the transfield equation itself.}}
By including the ``centrifugal'' term we simply modify the above 
conclusions and the transfield equation reduces to equation (\ref{hyperregime}).
\\
Thus, for $\sigma > \sigma_{c}$
the solution of the transfield equation is simply a straight line
$\zeta \approx (r_{s}/\varpi_0) \left[1+(\sigma_{s}/\tau)
(\varpi-\varpi_{s})/\varpi_{s}\right] $.
By employing equation (\ref{ode_cyl1}) we finally get
\begin{equation}\label{regime2}
z = r_{s}\Phi \left(1+\frac{\sigma_{s}}{\tau} 
\frac{\varpi-\varpi_{s}}{\varpi_{s}}\right)\,,
\quad
\sigma = \sigma_{s} \frac{\varpi}{\varpi_{s} }\frac{\Phi r_{s}}{z}\,.
\end{equation}
(In the most general, i.e., non-self-similar case, 
the solution is the conical Ia shape that we discuss
in \S~\ref{genanalysis}.)\\
Equation (\ref{jparallel}) gives
$J_\parallel \approx - \frac{c B_\phi \sigma}{4 \pi} \frac{F+2}{\tau}$.
Thus, for $F>-2$ the current density component $J_\parallel$ 
becomes positive at large distances.
The combination $-2<F<-1$ corresponds to a transition from current-carrying 
($J_\parallel <0$, at $\sigma \gg 1$)
to a return-current regime ($J_\parallel >0$ at $\sigma \ll 1$),
required by the current-closure condition.
\\
The azimuthal centrifugal force becomes important after the point where the two terms on the 
right-hand side of equation (\ref{trans1}) become comparable.
Using the scalings of equation (\ref{regime2}), this happens at
$\tau \mu^2 \approx x^2 \sigma^2$, 
so $\sigma_{c} \approx c \mu \sqrt{\tau} / \varpi_{mf} \Omega$.
\\
If we continue the integration for sufficiently large $z$, the fast decrease of $\sigma $ stops at
the point where $\sigma \approx \tau$. At this point the
product $B_p \varpi^2$ is of the order of $(-F) A$ (see eq. [\ref{flux2}]).
As we discuss in \S~\ref{genanalysis} the flow could continue to be accelerated after that point,
but with a much lower (logarithmic) rate.

\subsubsection{Numerical integration}
Numerically we may integrate the equations (\ref{ode_cyl}) as follows.
Suppose that we examine the flow along the field-streamline $A$ such that $\Phi(A)=1$,
and $\mu=\mu_0=10^6$, $\Omega = 100 $\ rad s$^{-1}$
(for which eq. [\ref{varpi_0}] gives the value of $\varpi_0$ $=c/\Omega =3 \times 10^8$ cm,
and the independent variable of the ODEs is $ \zeta=z \Omega/c$).
We give the model parameters ($\tau$, $F$), and a trial value for the 
$\sigma$ function at the modified fast-magnetosound singular point, 
$\sigma=\sigma_{mf}$.
At this point, both the numerator and denominator of equation (\ref{ode_cyl2}) vanish,
and we are able to find the values $ \zeta_{mf}$ and $\varpi_{mf}$. 
Using l'H${\hat {\rm o}}$pital's rule
we find the slope $\left(d \sigma / d  \zeta \right)_{mf}$ and start the integration
from the singular point downstream.
At some value of the independent variable 
$ \zeta= \zeta_{s}$ we find $\sigma = \sigma_{s}$ $(=0.003)$.
At this point the distance should be $z=r_{s}$ ($=3 \times 10^{17}$ cm),
and according to equation (\ref{z_cyl})
the equality $ \zeta_{s}=r_{s}/\varpi_0$ $(=10^9)$ should also hold.
If the latter equality is not satisfied, we change the trial value $\sigma_{mf}$ and
repeat the whole procedure until we find $\zeta_{s}=10^9$.
For the correct $\sigma_{mf}$ we are able to integrate downstream until 
$\sigma \approx \sigma_{s}$ 
(at the distance $r_{s}$, just before the termination shock), as well as upstream
until $\sigma = \sigma_{f} \approx 10^4$, i.e., until the solution encounters the 
classical fast-magnetosound surface.
(Using the value of the azimuthal magnetic field $B_{\phi,{s}}=-2 \times 10^{-5}$G
and eq. [\ref{bifield}],
we find the constant $B_0$.)

\qquad \underline{
Solution $a$}\\
The representative solution $a$ corresponds to the set of parameters
($\tau=10^{-5}\,, F = -1.9$). For $\sigma_{mf}=0.9$
the integration gives $\sigma_{s}=0.003$ at $r_{s}=3 \times 10^{17}$cm.
Figure \ref{fig1} shows the field lines as well as the current lines
in the poloidal plane.
The line shape is quasi-conical with small opening angle consistent with the model assumption
$B_z \gg B_\varpi$.
The current density component $J_\parallel \approx J_z$ 
changes sign from negative close to the classical
fast-magnetosound surface to positive at larger distances, satisfying (at least partly)
the current-closure condition
(according to the analysis of \S \ref{scalings}, this is expected for $-2<F<-1$).
\\
Figure \ref{fig2} shows the function $\sigma $ and the transition from Poynting- ($\sigma\approx 10^4$)
to matter- ($\sigma=0.003$) dominated regime.
The $\sigma$ function scales as $\sim z^F=z^{-1.6}$ at $\sigma>$ a few (force-free regime)
and as $\sim 1/z$ when the flow becomes matter-dominated.
\\
The components of the magnetic field are shown in figure \ref{fig3}.
The main component is the azimuthal one, and at the distance $r_{s}$ it becomes $2\times 10^{-5}$G.
It is evident from the figure that the self-consistency condition $B_z \gg B_\varpi$
is everywhere satisfied.
\\
Figure \ref{fig4} shows the transfield components of the various force densities as
functions of $z/r_{s}$ along the reference field line $\Phi=1$.
As expected, the dominant forces are ${f}_{B \bot}\approx J_\parallel B_\phi / c$ and 
${f}_{E \bot}= J^0 E /c$, and they almost cancel each other.
For $z/r_{s} < 2 \times 10^{-2}$, ${f}_{B \bot}>0$ 
and ${f}_{E \bot}<0$, corresponding to $J_\parallel <0$ and $J^0 <0$.
For larger $z/r_{s}$ both
${f}_{B \bot}$ and ${f}_{E \bot}$
change sign and the flow enters the return-current regime $J_\parallel >0 $,
where the charge density is positive ($J^0/c >0$).\\
The forces ${f}_{I \bot} + {f}_{EM2}$, ${f}_{EM1}$, and  ${f}_{C \bot}+{f}_{EM3}$
are proportional to the first, second, and third terms
in equation (\ref{transfieldfinal}), respectively.
At small heights ${f}_{I \bot} + {f}_{EM2} < 0$ and the curvature radius is ${\cal R} > 0$:
its exact value is controlled by the electromagnetic forces
$| {f}_{I \bot} + {f}_{EM2} | \approx  | {f}_{EM1} |$.
However, at $z/r_{s}=0.4$, we encounter 
the point $\sigma=\sigma_{c}\approx 8 \times 10^{-3}$,
where the ``centrifugal'' force $-{f}_{C \bot}-{f}_{EM3}$ (that mainly consist of
the azimuthal centrifugal part ${f}_{C \bot}$) becomes important. 
After that point ${\cal R}<0$ and the transfield equation
becomes $| {f}_{I \bot} + {f}_{EM2} | \approx  -{f}_{C \bot}$.
The small difference between these two terms is the electromagnetic term $| {f}_{EM1} |$.

\begin{figure*}[t]
\centerline{ {\includegraphics[ scale=.93]{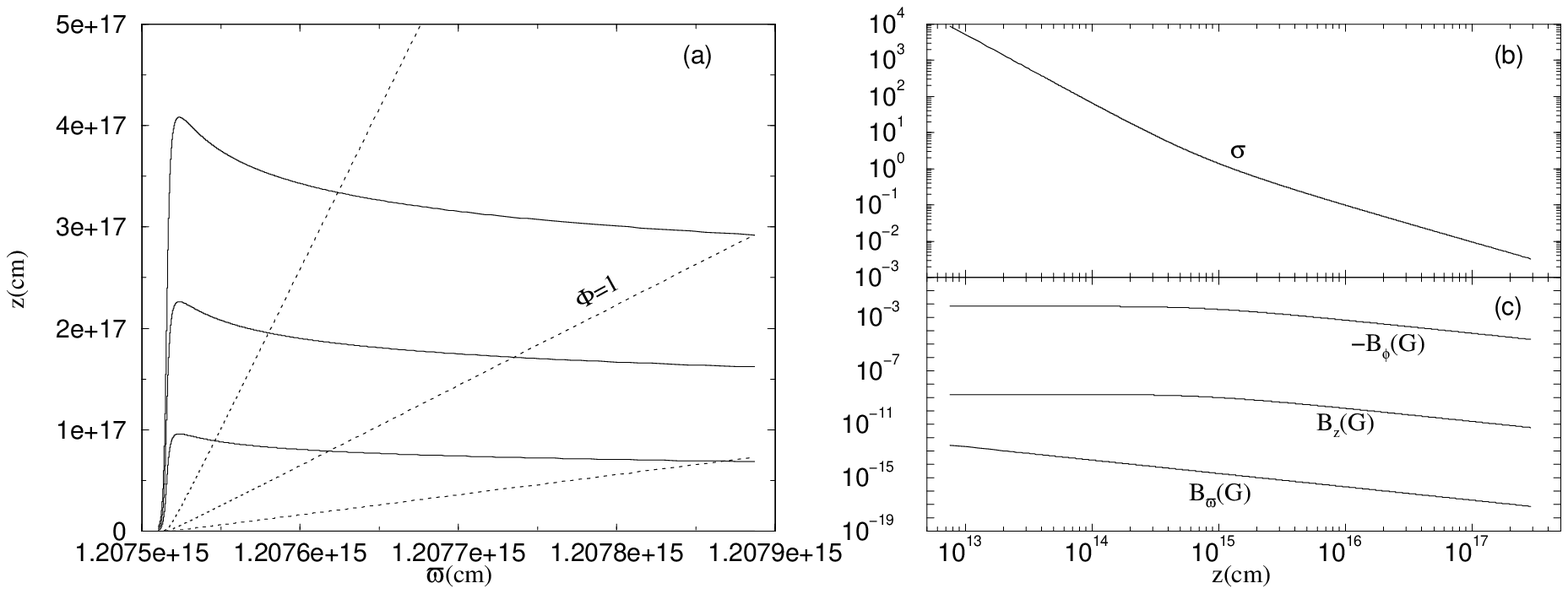}}}
  \caption{Solution $b$, corresponding to 
$\tau=10^{-6}$, and $F=-1.9$.
($a$) Three field lines ($\Phi=0.25\,, 1\,, 4$ -- {\emph {dashed}}) 
and three current lines ({\emph {solid}}).
($b$)--($c$) The Poynting-to-matter energy flux ratio $\sigma$
and the magnetic field components for the reference field line $\Phi=1$.
\label{fig5}}
\end{figure*}

\qquad \underline{
Solution $b$}\\
Figure \ref{fig5} shows another solution similar to solution $a$. The only difference is that here
$\tau=10^{-6}$.
Equations (\ref{ode_cyl}) show that 
for different $\tau$, the value $\sigma_{mf}$ remains roughly
the same, while $\varpi_{mf}$ 
scales as $(\varpi_{mf})_b =(\varpi_{mf})_a \times  \tau_a / \tau_b$.
Thus, the cylindrical distances in solution $b$ are ten times larger
compared to the ones for solution $a$.

\begin{figure*}
\centerline{ {\includegraphics[ scale=.93]{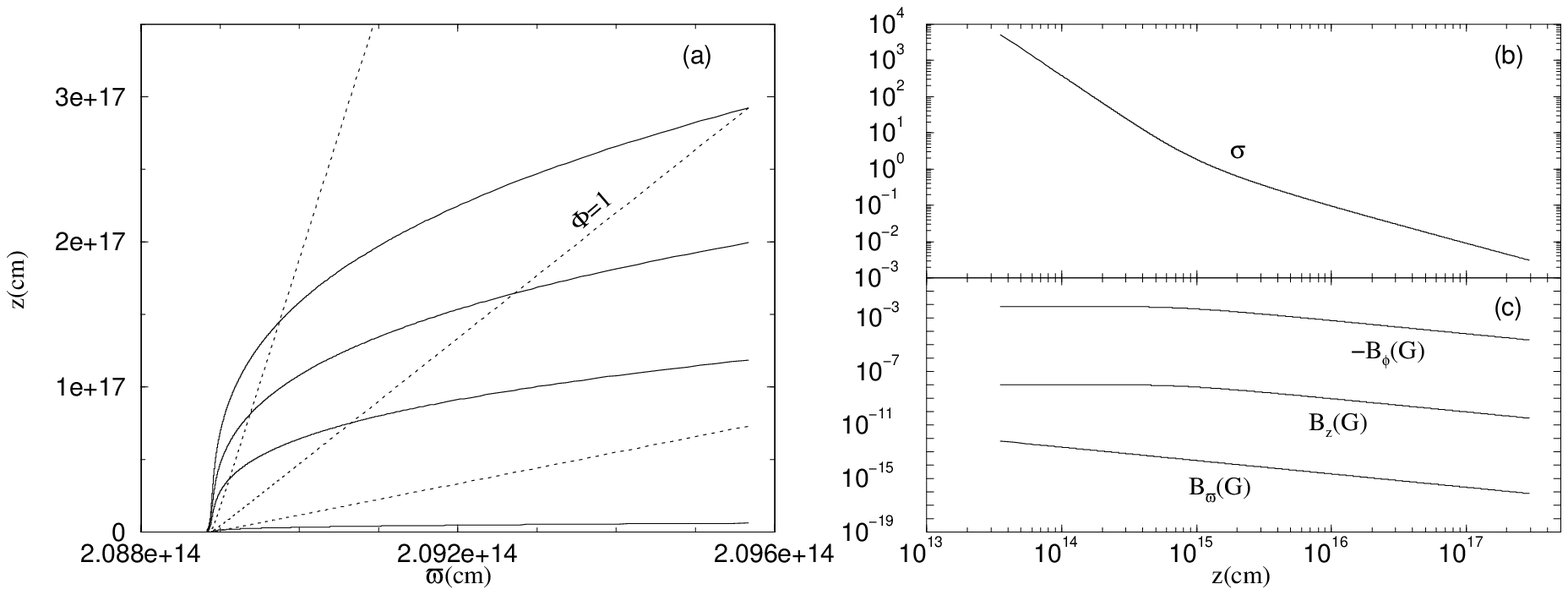}}}
  \caption{Same as fig. \ref{fig5}, but for the solution $c$, corresponding to
$\tau=10^{-5}$, and $F=-2.5$.
Here $J_\parallel <0$ all the way from the origin
to infinity, and there is no distributed return-current regime.
\label{fig6}}
\end{figure*}

\qquad \underline{
Solution $c$}\\
Figure \ref{fig6} shows a solution corresponding to $(\tau=10^{-6}\,, F=-2.5)$.
In this case the constrain on $\sigma$ at $r_{s}$ implies 
$\sigma_{mf}=1.5$.
Because $F<-2$, the current density component $J_\parallel$ remains negative
all the way from the source to infinity.
If we continue the integration to sufficiently large distances (not shown in fig. \ref{fig6})
the poloidal current becomes parallel to the flow and the acceleration stops; 
the function $\sigma $ then approaches a finite asymptotic value $\sigma_\infty (A)$.

\section{Asymptotic structure of efficient accelerators}\label{genanalysis}
The outcome of the magnetic acceleration is either $\sigma_\infty=\sigma_\infty(A)$
or $\sigma_\infty=0$.
Although the second possibility incorporates the current-closure condition
and seems more natural, the first one is also possible
(it requires that the current closes in a thin sheet at the end of the 
ideal MHD asymptotic regime, but it cannot be ruled out, simply because
all MHD winds practically terminate at some finite distance).

\begin{center}
\epsscale{1.1}
\plotone{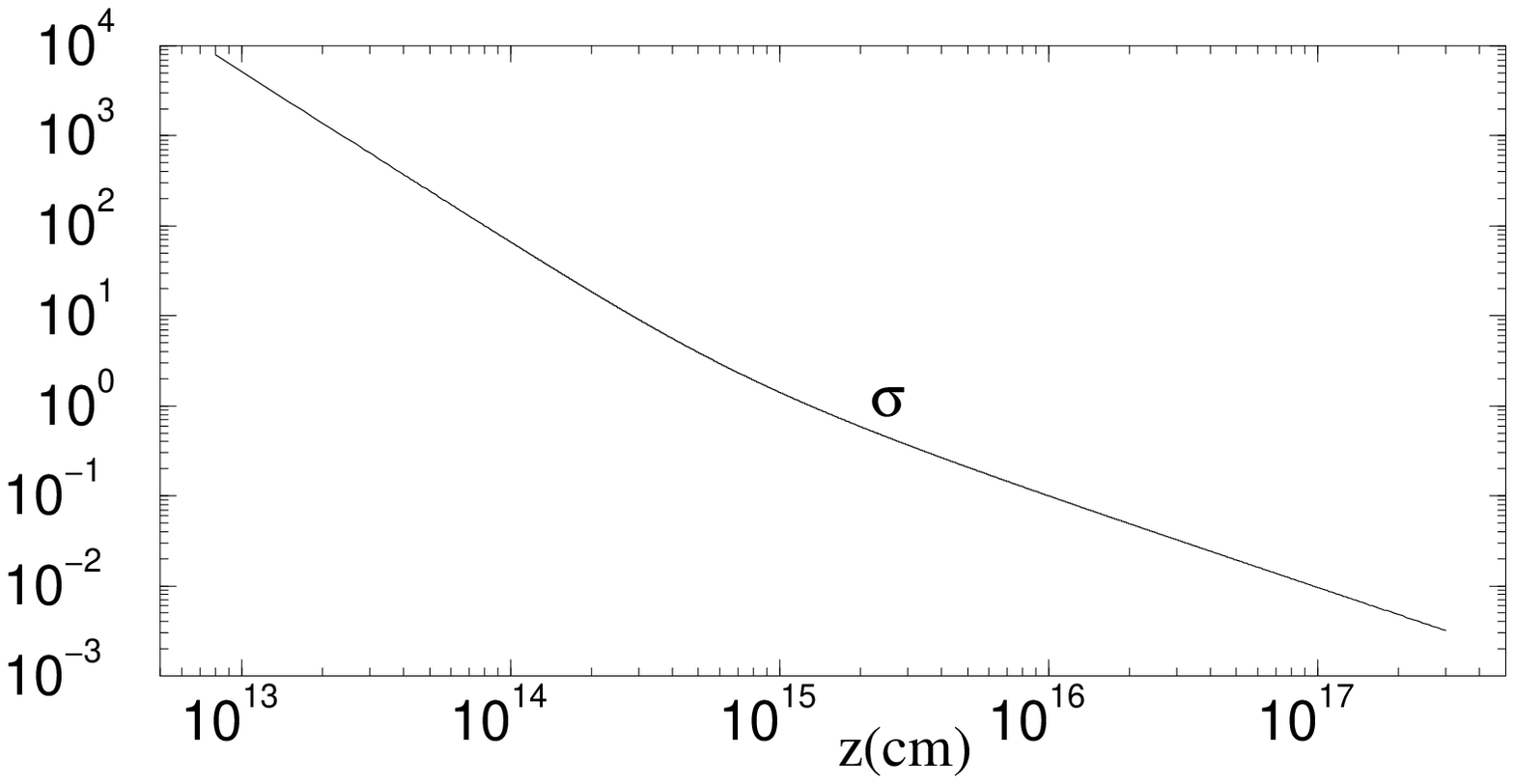}
\figcaption[]
{The Poynting-to-matter energy flux ratio $\sigma $ as a function of 
$z=\zeta c /\Omega$, along the reference field line $\Phi=1$ of solution $a$. 
\label{fig2}}
\end{center}
\begin{center}
\epsscale{1.1}
\plotone{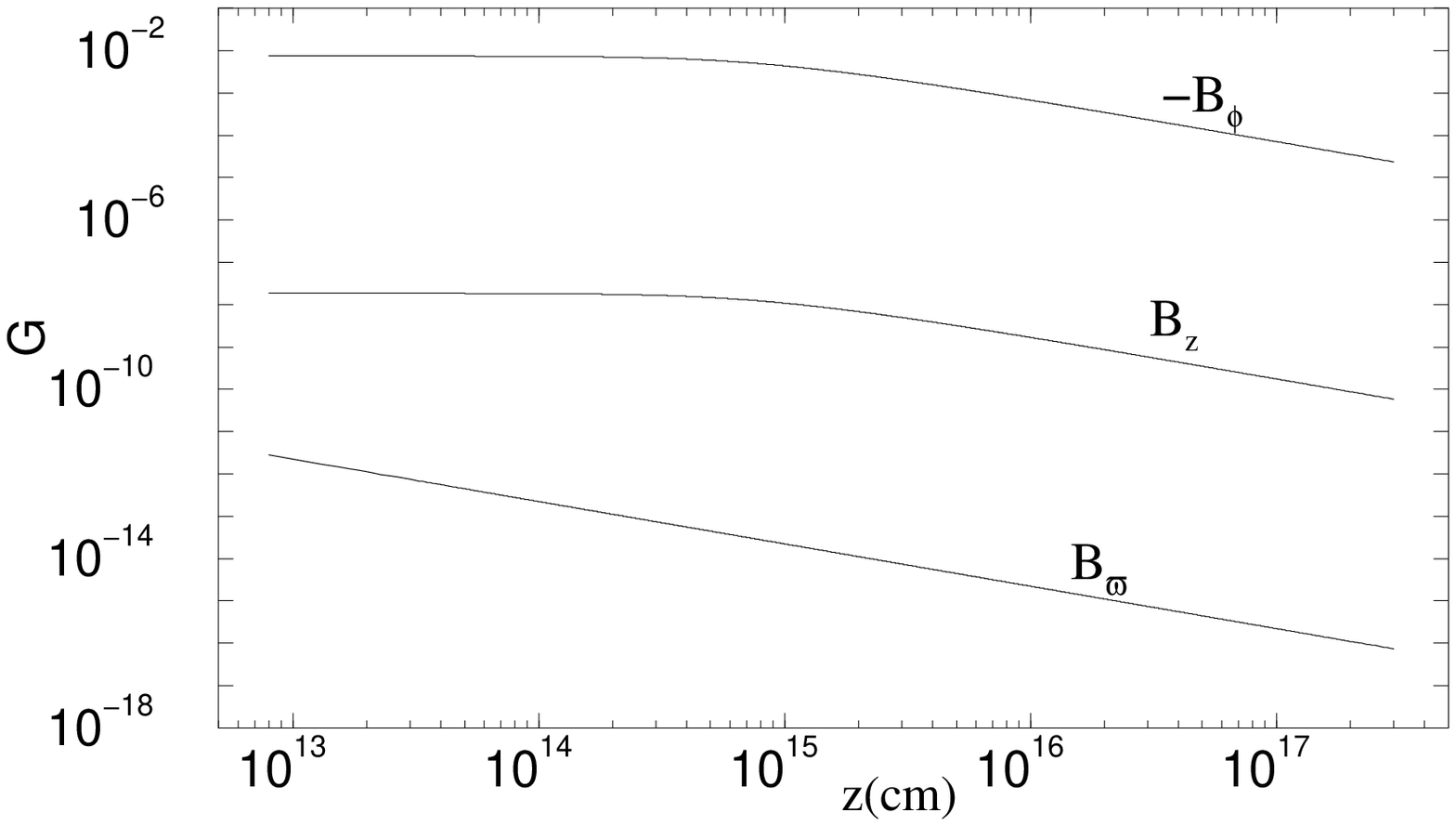}
\figcaption[]
{The magnetic field components as functions of $z$, 
along the reference field line $\Phi=1$ of solution $a$.
\label{fig3}}
\end{center}
\begin{center}
\epsscale{1.}
\plotone{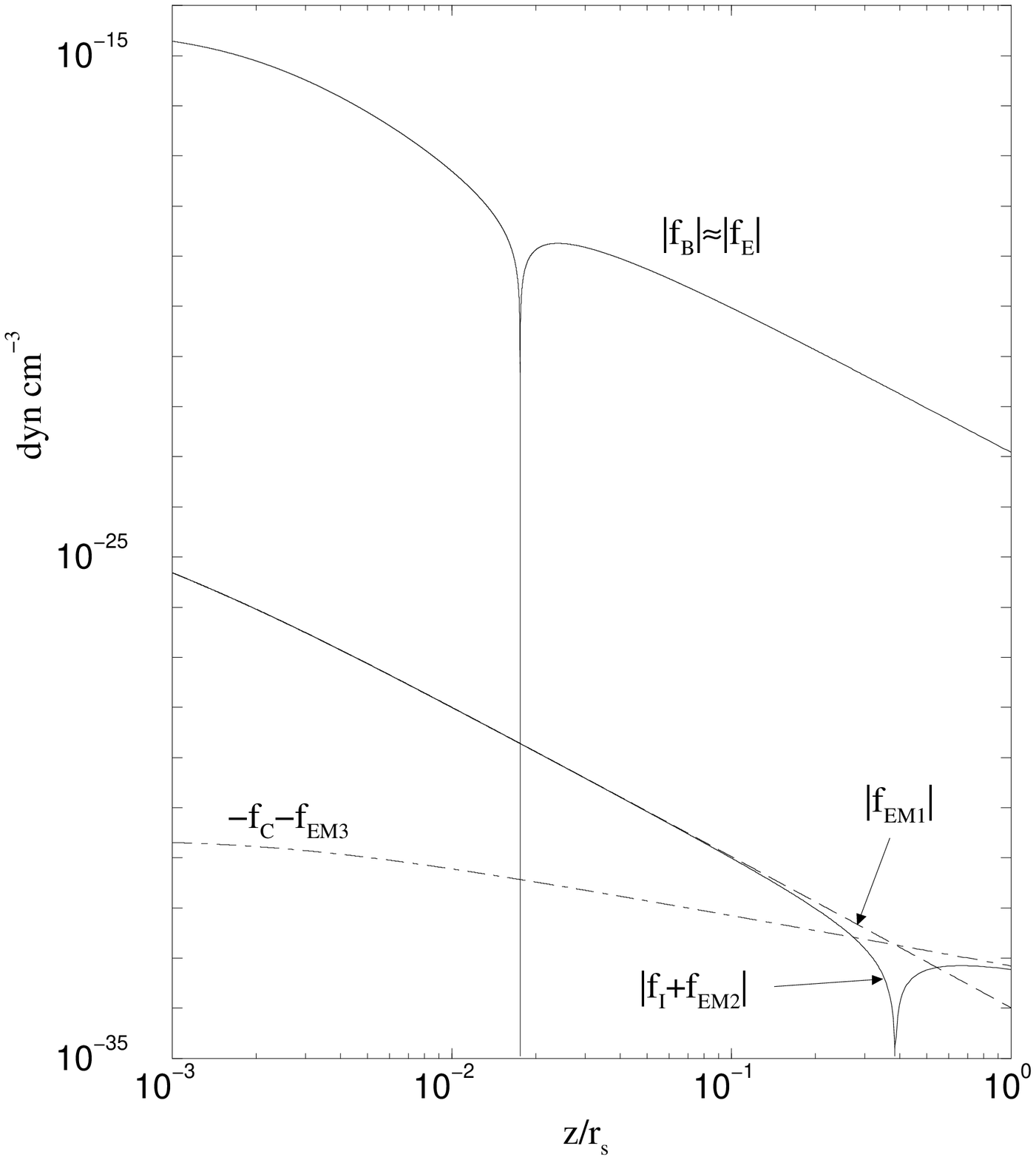}
\figcaption[]
{Transfield components of the various force densities
as functions of $z/r_{s}$, along the reference field line $\Phi=1$ of solution $a$.
The transitions from $J_\parallel <0$ to 
$J_\parallel >0$ (at $z/r_{s} = 2 \times 10^{-2}$, where $f_{B \bot}$ vanishes) 
as well as from ${\cal R} > 0$ to ${\cal R}< 0$ (at $z/r_{s} = 0.4$,
where $f_{I \bot}+f_{EM2}$ vanishes) are shown.
\label{fig4}}
\end{center}
Besides the value of $\sigma_\infty$,
another important issue is the acceleration rate.
In this section we focus on the field line asymptotic shape 
in the regime $\sigma<1$ and its relation to the acceleration.
(The analysis holds for nonrelativistic flows as well.)
Note that the discussion of the present section  -- as well as the characteristics of the
flow described in \S~\ref{character} -- is general and not restricted 
to the self-similar solutions presented in \S~\ref{zss}.
It applies in the asymptotic regime $\sigma < 1$
where the motion is mainly poloidal, $V_\phi \ll V_p$. 
Equation (\ref{omega}) yields $-\varpi B_\phi \propto \varpi^2 B_p $,
or, (using eq. [\ref{currdistr}]), $\sigma / (1 + \sigma) \propto \varpi^2 B_p$
(see also eq. [\ref{bernoullifinal}]).\footnote{In nonrelativistic
flows, for $V_p$ approaching its maximum value 
$V_{p \infty} \gg V_\phi$, a similar relation, namely
$-\varpi B_\phi \approx \varpi^2 B_p \Omega / V_{p \infty} $, holds.}
Thus, an accelerating flow corresponds to a decreasing $\varpi^2 B_p$,
a function that depends on the line shape.
In other words, how fast the acceleration takes place
depends on the magnetic flux distribution,
by the solution of the transfield equation. 

We already know that a field line shape of the form $z=c_1(A) \varpi^{c_2(A)}$ gives only
logarithmic acceleration (Chiueh et al. 1991).
The deviations from monopole magnetic field also
result in inefficient \citep{BKR98}, or logarithmic \citep{LE01} acceleration. 
(The monopole geometry itself
-- which however does not satisfy the transfield equation --
gives extremely inefficient 
acceleration: $\sigma_\infty =\mu / \gamma_\infty= \mu^{2/3}\gg 1$; \citealp{M69}.)
The logarithmic acceleration gives asymptotically $\sigma_\infty=0$,
but over completely unrealistic (exponentially large) distances.
The numerical simulations by \citet{B01} also give inefficient acceleration, possibly because the
initial configuration is a split monopole field
and does not change much during the simulation.

The $r$ self-similar solutions of Li et al. (1992) and \citet{VK03a}
give final equipartition $\sigma_\infty =1$, but the same model could give smaller
$\sigma_\infty $ as well \citep{VK03b,VPK03}.
However, because it cannot capture the transition from positive to negative curvature
radius, it is impossible to obtain $\sigma_\infty=0$.
The asymptotics of the $r$ self-similar model are straight poloidal field lines
($\varpi/{\cal R}=0$); as a result,
the transfield component of the electromagnetic force 
asymptotically equals the ``centrifugal'' force (which could be very small, but definitely
not exactly zero), and $\sigma_\infty=\sigma_{c}$.
Even so, the value $\sigma_{c}$ could be very small, and the model should be explored more
carefully in that respect.

In the $z$ self-similar model presented in \S~\ref{zss}, we kept the line shape free:
$z\propto \Phi(A) \zeta(\varpi)$, and the solution itself finds the shape through
the function $\zeta(\varpi)$.
The result is a quasi-conical shape, and the distribution of the magnetic flux
is different from the monopole field (see eq. [\ref{flux}]).

\begin{table*} 
\begin{center}
\caption{
Possible asymptotic geometries and the corresponding acceleration$^\dagger$
\label{table1}}
{
\small
\begin{tabular}{ccccc}
\tableline
\tableline
& \multicolumn{4}{c}
{The far-asymptotic regime (after the end of the main acceleration)} 
\\
\cline{2-5}
&
$\sigma \rightarrow 0$
&
$0<\sigma_\infty<\sigma_{c}$
&
$\sigma_\infty=\sigma_{c}$
&
$\sigma_\infty>\sigma_{c}$
\\
&
$I \rightarrow 0$
&
$I_\infty=I_\infty(A) \neq 0$
&
$I_\infty=I_\infty(A) \neq 0$
&
$I_\infty=I_\infty(A) \propto \gamma_\infty(A)^*$
\\
&
$f_{I \bot} =- f_{C \bot}$
&
$(f_{I \bot} =- f_{C \bot})_\infty$ 
&
$(f_{EM1}  =- f_{C \bot})_\infty$
&
$(f_{EM1})_\infty  =0$
\\
&
parabolic II or IIa
&
hyperbolic 
&
conical I or Ia
&
conical I or Ia
\\
The $\sigma_{c} < \sigma < 1$ acceleration phase
&
&
&
or parabolic I or Ia
&
or parabolic I or Ia
\\
$f_{EM1}  =- f_{I \bot} $
&
&
&
or cylindrical
&
or cylindrical
\\
\tableline \tableline
conical Ia:
$\ z=z_0(A) + {\varpi}/{\tan \vartheta(A)}$
& 
&
&
&
\\
$1/\varpi^2 B_p =
{\vartheta^{'} / \sin \vartheta - z_0^{'} \sin \vartheta / \varpi}$
& 
$\surd$
&
$\surd$
&
$\surd$
&
$\surd$
\\
\tableline
parabolic Ia:
$\ z=z_0(A) + c_1(A) \varpi^{c_2}$, $c_2=$const
& 
&
&
&
\\
$1/\varpi^2 B_z =
{(-c_1^{'}/c_1 c_2) - ( z_0^{'} / c_1 c_2) \varpi^{-c_2}}$
& 
$\surd$
&
$\surd$
&
$\surd$
&
$\surd$
\\
\tableline
parabolic II:
$\ z=c_1(A) \varpi^{c_2(A)}$
& 
&
&
&
\\
$1/\varpi^2 B_z =
{(-c_1^{'}/c_1 c_2) - (c_2^{'}/c_2) \ln \varpi }$
& 
$\surd$
&
$\surd$
&
---
&
---
\\
\tableline
parabolic IIa:
$\ z=z_0(A)+c_1(A) \varpi^{c_2(A)}$
& 
&
&
&
\\
$1/\varpi^2 B_z =
{(-c_1^{'}/c_1 c_2) - (c_2^{'}/c_2) \ln \varpi -( z_0^{'} / c_1 c_2) \varpi^{-c_2}}$
& 
$\surd$
&
$\surd$
&
---
&
---
\\
\tableline
\multicolumn{5}{l}{$^\dagger$ The table covers the asymptotics of nonrelativistic flows as well.}
\\
\multicolumn{5}{l}{$^*$ This is the solvability condition at infinity \citep{HN89,CLB91}.}
\\
\end{tabular}
}
\end{center}
\end{table*}

Collimation of relativistic outflows is in general much more difficult than
in the nonrelativistic cases, firstly because the electric force almost cancel
the transfield component of the magnetic force,
and secondly because the effective matter inertia is larger (e.g., \citealt{B01}).
The result is negligible curvature, $\varpi /{\cal R} \approx 0$ (e.g., \citealt*{CLB98}),
and only close to the origin where the Lorentz factor is smaller than a few tens
is collimation efficient \citep{VK03b}.
This result, together with the envisioned close to $z = \varpi / \tan \vartheta (A)$
field line shape
-- we employ the term ``type I conical'' for this shape --
led to the erroneous conclusion that acceleration
is impossible when $\varpi /{\cal R} \approx 0$.
However, negligible curvature does not necessarily mean $z \propto \varpi$;
the most general case is a ``type Ia conical'' shape given by
\begin{equation}\label{straightshape}
z=z_0(A) + \frac{\varpi}{\tan \vartheta (A)}, \quad z_0^{'} \neq 0\,. 
\end{equation}
Differentiating equation (\ref{straightshape}) we get 
for the poloidal magnetic field
\begin{equation}\label{straightfield}
\varpi^2 {\boldsymbol {B}}_p = \varpi {\boldsymbol{\nabla}} A \times \hat \phi =
\frac{\cos \vartheta  \  \hat z + \sin \vartheta  \ \hat \varpi}
{\vartheta^{'} / \sin \vartheta - z_0^{'} \sin \vartheta / \varpi} \,.
\end{equation}
(Here primes denote derivative with respect to $A$.)
It is evident from equation (\ref{straightfield}) that 
for $z_0^{'} > 0$, $\vartheta^{'} > 0$ the quantity
$\varpi^2 B_p$ is a decreasing function of $\varpi$, meaning that
the flow is accelerated. 
Magnetic flux conservation implies $2 \pi \delta A = B_p \delta S$, where
$\delta S$ is the cross-section area between two neighboring magnetic flux surfaces
$A$, $A+\delta A$ ($\delta S = 2 \pi \varpi \delta \ell_\bot$, see fig. \ref{fig7}). 
Since $B_p \varpi^2 / \delta A = 2 \pi \varpi^2 / \delta S$,
it is clear that acceleration is possible when -- as the flow moves --
$\delta S$ increases faster than $\varpi^2$. This is the case in type Ia conical flow,
while in type I conical it is exactly $\delta S \propto \varpi^2 $
and the quantity $B_p \varpi^2$ remains constant.
The previous analysis shows that the argument of Chiueh et al. (1998)
related to the global inefficiency of the 
magnetic acceleration in relativistic flows (based on the fact that 
the collimation is inefficient), can be circumvented.
The conical Ia field lines are perfectly straight, satisfying the
$\varpi / {\cal R} =0$, nevertheless, they expand in the sense
that $\delta S$ increases faster than $\varpi^2$.
In other words, we may have expansion (increasing $\delta S/\varpi^2$), 
without bending ($\varpi / {\cal R} \neq 0$).
\\
By combining equations (\ref{straightfield}) and (\ref{bernoullifinal}), we find
that the $\sigma $ function could change from a $\sim 1$ value at 
$\varpi \sim A z_0^{'} \sin \vartheta / (A \vartheta^{'} / \sin \vartheta - 
2 \sigma_{\rm M}/\mu)$
(the conical Ia shape starts roughly at this distance), and becomes
$\sigma=\sigma_{\rm min}$
at distances $\varpi \gg z_0^{'} \sin^2 \vartheta /\vartheta^{'}$, with
\begin{equation}
\label{sigmamin}
\sigma_{\rm min}=\frac{\sigma_{\rm M}}{\mu} \left(\frac{B_p \varpi^2}{A}\right)_{\rm min}=
\frac{\sigma_{\rm M}}{\mu} \frac{\sin \vartheta }{ A \vartheta^{'}}\,.
\end{equation}
Since the factor $\sin \vartheta /A \vartheta^{'}$ is not $\ll 1$ (this would require the
opening of the field lines asymptotically to change rapidly from 
line-to-line which is unlikely to happen)
the only case that gives $\sigma_{\rm min} \ll 1$ is the $\mu \gg \sigma_{\rm M}$.
This is an important requirement for the above mechanism to work,
and it is equivalent to $B_p \varpi^2 \gg A$ 
in the super-Alfv\'enic regime and as long as the $\sigma$ is $\gg 1$
[in this regime $\mu \approx - \varpi \Omega B_\phi/ \Psi_A c^2 \approx
B_p \varpi^2 \Omega/ \Psi_A c^3 = (B_p \varpi^2 / A) \sigma_{\rm M} $].
Hence, as \citet{CLB98} state,
the poloidal field lines need to be bunched within a small
solid angle before expansion takes place.
Contrary to their claim, however, the expansion is not
necessarily sudden near the light cylinder, but extended to $x \gg 1$.
In \S~\ref{equatorialwind} we discuss how such a field configuration may be created
(see also fig. \ref{fig7}).

The general asymptotic expansion of the field line configuration -- in both, relativistic and
nonrelativistic cases -- is
\begin{equation}
z=z_0(A) + c_1(A) \varpi^{c_2(A)}\,,
\end{equation}
for which we get
\begin{equation}
\varpi^2 B_z = \varpi \frac{\partial A}{\partial \varpi} = 
\frac{c_2}{-c_1^{'}/c_1 - c_2^{'} \ln \varpi -( z_0^{'} / c_1) \varpi^{-c_2}}\,.
\end{equation}
Chiueh et al. (1991) examined two types of parabolic cases: 
type I ($z_0^{'}=0\,, c_2\neq 1\,, c_2^{'}=0$) when $\varpi^2 B_z$ is a field line constant, and
type II ($z_0^{'}=0\,, c_2\neq 1\,, c_2^{'}\neq 0$) when 
the $\varpi^2 B_z $ slowly (logarithmically) decreases and asymptotically vanishes.
They also examined the conical type I case ($z_0^{'}=0\,, c_2=1$).
Besides the conical type Ia case,
new types of parabolic cases corresponding to $z_0^{'} \neq 0$
should be added in the analysis.
We henceforth call them type Ia, for $c_2=$const$\neq 1$,
and type IIa, for $c_2=c_2(A)\neq 1$.
These new parabolic types together with the type Ia conical, 
are much faster than logarithmic accelerators in
relativistic as well as in nonrelativistic flows.
Table \ref{table1} summarizes the various possible geometries in the
acceleration phase $\sigma_{c}< \sigma<1$ (first column) combined with
the possible final results of the acceleration -- including possible
acceleration after the $\sigma=\sigma_{c}$ transition --
and the far-asymptotic final geometries. 
The final current distribution and the dominant forces in the transfield 
direction are also shown.
Compatible combinations of the geometry in the acceleration phase
$\sigma_{c}< \sigma<1$
and the far-asymptotic final state are marked with a ``$\surd$''.
For example, if the flow has a type Ia conical shape
(this is the most plausible case for relativistic MHD in which
the field lines have negligible curvature $\varpi/ {\cal R} \ll 1$),
the $\sigma$ function decreases according to equation (\ref{bernoullifinal})
and the shown in the Table \ref{table1} function $\varpi^2 B_p$, and 
reaches a minimum value $\sigma_{\rm min}$ at distances 
$\varpi \gg z_0^{'} \sin^2 \vartheta / \vartheta^{'}$ (see eq. [\ref{sigmamin}]).
If the azimuthal centrifugal force is still negligible at these distances
(i.e., $\sigma_{\rm min}>\sigma_{c}$) the acceleration stops
and the far-asymptotic regime is the one in the fifth column (most likely
with unchanged conical Ia shape).
If $\sigma_{\rm min} \approx \sigma_{c}$, then the azimuthal centrifugal force
starts to play a role in determining the poloidal curvature.
One possibility is that the acceleration does not continue even with
slightly different line shape, resulting in the fourth column case.
The remaining case is that the $\sigma$ continues to decrease, and the 
line shape becomes hyperbolic.
Since the asymptotic form of a hyperbolic flow is again conical,
the $\sigma $ function reaches a new minimum given from an equation 
similar to (\ref{sigmamin}). 
If the acceleration stops at this point, we get a
final situation corresponding to the third column of Table \ref{table1}.
However, even tiny deviations from an {\emph{exactly}} radial flow 
$\varpi / {\cal R}=0$ could give a slow (logarithmic) acceleration to the 
completely matter-dominated regime $\sigma_\infty=0$, as was argued by \citet{O02}.
This situation corresponds to a continuous acceleration under a
parabolic II or IIa shape (second column).
We note, however, that for realistic applications,
the interaction of an outflow with its environment
brakes-down the ideal MHD conditions and this probably happens before
the realization of this last part of the acceleration.

Chiueh et al. (1991) examined the far-asymptotics of relativistic MHD outflows,
generalizing the work on nonrelativistic flows by \citet{HN89}. 
They assumed that the various quantities
become functions of $A$ alone (the characteristic of the far-asymptotic regime),
neglected the poloidal centrifugal term ${\cal O} (\varpi / {\cal R})$ as well as 
terms of order ${\cal O} (x_{\rm A}^2/x^2)$ in the transfield equation,
and found that conical field lines satisfying $(z/\varpi)_\infty < \infty$
should enclose a finite current $I_\infty = I_\infty(A) \propto \gamma_\infty(A)$
(the solvability condition at infinity, see \S \ref{character}).
As already discussed in \S \ref{character}, this is the case when
$\sigma_\infty > \sigma_{c}$, but
for smaller values of $\sigma_\infty$ the 
asymptotic transfield equation reduces to
the balance between the poloidal and azimuthal centrifugal forces
[i.e., they are exactly the ${\cal O} (x_{\rm A}^2/x^2), {\cal O} (\varpi / {\cal R})$ terms that remain
in the transfield];
the electromagnetic term is much smaller, but nevertheless not exactly zero and no
solvability condition can be derived.
Our self-similar solutions fall in this category:
the condition $I_\infty(A) \propto \gamma_\infty(A)$
is not satisfied, because $\sigma_\infty < \sigma_{c}$.

As we see in Table \ref{table1} the solvability condition at infinity
is mandatory only for the fifth column. 
As a result we may have, e.g., nonrelativistic conical
or type I/Ia parabolic flows that carry poloidal current $I_\infty=I_\infty(A)$
(with negative or positive $dI_\infty/dA$ when they are in the 
current-carrying or return-current regime respectively), provided that
$\sigma_\infty \leq \sigma_{c}$.

As we already stated, the analysis of this section applies to nonrelativistic flows as well.
A first step towards finding nonrelativistic MHD solutions with $\sigma_\infty \approx 0$ is the
super-- modified fast-magnetosound solution of \citet{V00}, which ends slightly
after the modified fast singular point with $\sigma = 0.005$.
The full acceleration ($\sigma_\infty=0$) 
should be examined using a model that allows a transition from
positive to negative curvature radius (the asymptotic transfield force-balance equation
for nonrelativistic flows implies that the curvature radius has the sign of the $-J_\parallel$,
thus changing sign when the flow moves from the current-carrying to the return-current regime,
e.g., \citealp{O01}).

\section{The pulsar magnetosphere/equatorial wind}\label{equatorialwind}
The most efficient acceleration mechanism that we described
is the one based on the conical Ia field line shape. 
The $z$ self-similar solutions are just a numerical
example confirming that the 
mechanism indeed works. Although the presented solutions
describe an outflow very close to the polar axis, the same exactly physical mechanism
is expected to work for the equatorial wind as well. 
Only practical difficulties do not allow us to
derive semianalytical solutions in that regime. 

As we discuss in \S~\ref{genanalysis}, 
an important requirement for the above mechanism to work is that $\mu \gg \sigma_{\rm M}$.
In fact, as we argue below, such a configuration is most likely to happen near the equator.

The pulsar magnetosphere has been traditionally considered as being force-free, and
inertial effects are usually neglected. On the other hand, it is generally accepted that it is 
because of the inertia that the magnetic field line topology becomes open
and an azimuthal magnetic field component is created. 
Moreover, the pair creation and their acceleration to high
Lorentz factors (e.g., \citealt{DH82}) involve non-ideal MHD effects
that may be important in the final form of the magnetosphere.
A question arises: is the force-free a good approximation?

\citet{CKF99} have solved the force-free problem (although with the current-closure
not fully satisfied) and \citet{OK03} have repeated their numerical work with
higher resolution, deriving similar results. 
The usual problem with the force-free dynamics is that the condition $B^2-E^2>0$ is not
always satisfied, since the Bernoulli equation that 
relates the electromagnetic field with the Lorentz factor of the
system of reference where the electric field vanishes, is omitted
(see eq. [\ref{bernoulli-comB}]). In regimes where
$B^2<E^2$ the drift velocity is larger than the light speed.
This is indeed the case in the aforementioned force-free solution.
According to \citet{OK03}
the drift velocity becomes larger than  the light speed at distance a few times
the light cylinder (their figure 5 shows that typically this happens at cylindrical distance
$\approx 3 c / \Omega$). 
Thus, the force-free assumption brakes-down much before the
fast-magnetosound point ($\sigma = \mu^{2/3}$).

Another problem of the force-free solution is that it implies negative azimuthal
velocities near the source. \citet{CKF99} demonstrated how we can find
the flow speed in cases where the electromagnetic field is known. 
We need to solve the system of equations (coming from
eq. [\ref{omega}], and the second of eqs. [\ref{V}])
\begin{mathletters}
\begin{eqnarray}
\frac{V_\phi}{cx}=1-\frac{V_p}{c} \frac{2 | I |}{\Omega B_p \varpi^2}\,, \label{forcefree1}
\\
\xi \gamma \left(1- x \frac{V_\phi}{c}\right)= \mu (1-x_{\rm A}^2)\,,\label{forcefree2}
\end{eqnarray}
\end{mathletters}
together with the identity $1-1/\gamma^2 = (V_p/c)^2+(V_\phi/c)^2$.
The solution for $\gamma$ is actually equation (\ref{bernoulli-comB}).\footnote{
Since the back-reaction of the matter to the field is ignored, it is possible to find
$1/\gamma = 0$. This happens exactly in the positions were the drift velocity
equals the light speed, i.e., where $B^2=E^2$.}
The problem related to negative azimuthal speeds comes from equation (\ref{forcefree1}).
The second part of the right-hand side is much smaller than unity 
very close to the source in a possible corotating regime, and should increase up to
an asymptotic value equal to unity as we
move to larger $x$ (in order to have a negligible
asymptotic azimuthal speed). Practically speaking, the ratio $2 | I | / \Omega B_p \varpi^2$
should be $\approx 1 $ at distances $x\gtrsim$ a few.
After that point $\left(B_p \varpi^2\right)_{x>\mbox{a few}} \approx 2 | I | / \Omega$,
and this continues to be the case even in the
superfast (non-force-free) regime where both the $| I |$ and the $B_p \varpi^2$ decrease.
But, what happens relatively close to the light cylinder? 
The poloidal field lines follow the dipolar shape up to the point where they
approach the equator, but then they remain parallel to it
for some distance. We know that a relativistic flow
is difficult to bend and this is a characteristic of a force-free regime as well.
The curvature $1/ {\cal R}$ is of the order of 
$(1/z) \times {\cal O}\left( {{\rm max} }\{1/x^2\,, 1/\gamma^2\}\right)$,
meaning that at distances $x\gtrsim$ a few, the poloidal field lines are
practically straight.\footnote{
The $z / {\cal R} = {\cal O}\left( {{\rm max} }\{1/x^2\,, 1/\gamma^2\}\right)$
holds if we assume that $\hat{n}\cdot {\boldsymbol{\nabla}} \sim 1/z$ 
in eq. (\ref{transfield}).
Very close to the midplane, the field may change in a small scale.
In addition, the pressure gradient term may be important.
It is because of these two effects that the field lines do not cross the equator.}
As a result, the quantity $\varpi^2 B_p$ increases, because the surface between
two neighboring field lines increases slower that $ \propto \varpi^2$.
For a force-free case where $| I |$ remains constant, we get a
decreasing ratio $(V_p / c) (2 | I | / \Omega B_p \varpi^2)$. 
Since its asymptotic value should be unity, near the source is larger than unity, 
implying $V_\phi < 0$. In fact \citet{CKF99} found negative azimuthal velocities. 
The only way to avoid this effect in the framework of force-free field, is to have very small
initial poloidal velocity ($V_p\ll c$), meaning that our ``initial'' surface is
very close to the pulsar. However, in that case we should take into 
account non-ideal effects related to the pair creation region.

A solution to this problem could be that the flow in the regime
where the lines remain approximately parallel to the equator is
{\emph {decelerating}}, meaning that kinetic (or enthalpy) energy flux
is transfered to Poynting, and the $\sigma$ function increases. In that case, the
$| I | = (c/2) \varpi | B_\phi| $ would increase and cancel the increase of
$B_p \varpi^2$ in the expression of $V_\phi$, resulting in $V_\phi>0$.
During the phase where the flow is approximately parallel to the equator,
the $B_p \varpi^2=\delta A (2 \pi \varpi^2)/\delta S $ increases almost linearly with $\varpi$
($\delta S \approx 2 \pi \varpi \delta z$ with $\delta z \approx const$). Since
the $\varpi | B_\phi |$ is forced to follow this increase (in order
to be consistent with a negligible azimuthal motion), the matter energy
part is decreasing linearly with $\varpi $, and an important (or even the most significant) part of the
Poynting flux may be created there.
As the flow decelerates, at some point it reaches $\gamma \sim 10$. Only after that point the 
curvature of the poloidal field lines $(1 / {\cal R})$ may be important and the lines start
to expand to larger heights above the equator. The expansion in principle allows for acceleration
(if $\delta S $ increases faster than $\varpi^2$),
and this time the electromagnetic energy is transfered to the matter. We consider this point
as the origin of the ideal MHD regime that we examine in this paper.

Another problem comes when we consider the ``last'' current loop ($I=0$)
that closes on the equator. Along this equatorial current line the azimuthal magnetic field
vanishes. Since $B^2-E^2= B_\phi^2 - B_p^2(x^2-1)$, the inequality $B^2-E^2>0$
holds only in the unlikely case where the poloidal field also vanishes on the equator. 
The remaining case is to include non-ideal MHD effects (then the electric field is
not equal to $x B_p$ and $ B^2-E^2 = B_p ^2- E^2$ could be positive).

Although the details of the full magnetosphere remain to be explored, 
and this requires to solve the 
full problem (not only the non-force-free, but also the non-ideal MHD especially near
the ``last'' current line that closes on the equator; not to mention the inner/outer gaps and the
non-axisymmetry), the picture described above looks appealing
and consistent with the observed Crab-pulsar equatorial wind.
The scenario also explains why the quantities  
$B_p \varpi^2/A$, $2 | I | / A \Omega \approx \mu / \sigma_{\rm M}$, and $1/\sigma$
are likely $\gg 1$ just after the end of this ``parallel to the equator'' phase.
These are the ``initial''  conditions for the ideal MHD phase.
It may not be just a coincidence that the efficient and faster than logarithmic
acceleration to low-$\sigma$ values during the ideal MHD
phase was possible for these ``initial'' conditions only.
Note also that the part of the total open magnetic flux that contributes
to the ``parallel to the equator'' phase, is $\sim 50\%$ 
(for exactly dipolar field, $45\%$ of the open magnetic flux correspond 
to field lines that have $B_z<0$ when cross the light cylinder).
The rest $\sim 50\%$ of the open field lines likely follows a dipolar shape
till $x \sim $ a few, and then becomes conical Ia (again $\varpi / {\cal R}$ 
is negligible). Since there is no obvious reason for an increasing
$B_p \varpi^2$, the acceleration gives asymptotically $\sigma_{\infty} \sim 1$ or $\sigma_i $,
if the initial $\sigma $ value is $\sigma_i \gg 1$ or $\sigma_i \ll 1$, 
respectively.\footnote{The creation of bunched poloidal field lines
near the polar region cannot be completely ruled out.
For example, the pressure of the created radiation may push the lines 
towards the axis. Another reason may be the Poynting flux 
$(c/4 \pi)E_\parallel B_\phi$ associated with the parallel electric field inside the
non-ideal MHD regime, which points toward the axis for $\theta \gtrsim 55 \degr$
where $E_\parallel < 0 $.}

The ``initial'' conditions $B_p \varpi^2 \gg A$ are in principle possible in other
astrophysical settings as well, such as gamma-ray burst sources or AGNs. For example,
a corona or a thermal/radiation
pressure field may keep the field lines that emanate from a disk parallel to each other
(at least for some distance), in
which case the product $B_p \varpi^2 $ again increases and a matter energy-to-Poynting flux 
transformation is realized (e.g., \citealp{VK03b,VPK03}).

\section{Summary}\label{conclusions}
In the first part of the paper (\S~\ref{governing}, \ref{superalfvenic})
we derived the ideal MHD equations that describe the super-Alfv\'enic asymptotic regime.
The resulting system of equations remains intractable enough, and we could only find
a $z$ self-similar model (\S~\ref{zss}) describing the regime near the rotation axis.
We regard the solution $a$ as the best representative result of that model.
It includes all the expected characteristics of an efficient MHD accelerator, in particular:
\begin{itemize}
\item
The transition from $\sigma= 10^4$ to $\sigma \approx 0$, showing that ideal
MHD can account for the full acceleration, resulting in a matter-dominated flow.
\item
The transition from a current-carrying to a return-current regime,
showing how the poloidal current lines close.
\item
The transition from positive to negative poloidal curvature radius.
When the flow becomes sufficiently matter-dominated ($\sigma< \sigma_{c}$; see \S \ref{character})
it has a negative curvature radius and the line shape becomes hyperbolic.
This is simply the result of angular momentum
conservation: the azimuthal component of the velocity decreases, and the constancy of
the total velocity implies an increasing cylindrical component.
(This small deviation from a conical line shape could be enhanced in the termination shock.)
The most important implication of this effect is not the hyperbolic line shape
itself (which still remains
practically conical), but the new status quo in the transfield force-balance.
It is now the poloidal and azimuthal centrifugal inertial forces that dominate and their difference
is the much smaller transfield electromagnetic force component.
\item
The transition from sub-- to super-- modified fast-magnetosound flow.
This surface should be crossed since in the matter-dominated regime
the fast-magnetosound speed becomes progressively smaller (it vanishes at $\sigma=0$).
In this sense it is a signature of an efficient acceleration.
It is also related to the causality principle: only in the super-- modified fast-magnetosound
regime none of the MHD waves can propagate upstream and reach the origin of the flow.
\end{itemize}
We tried to present our numerical solutions in accord with the model of Kennel \& Coroniti
for the Crab nebula.
In solution $a$ the transition from $\sigma =10^4$ to $\sigma_{s}=0.003$ happens over 
less than five decades in distance from the pulsar,
and the value $\sigma_{s}$ is reached at distance of $3\times 10^{17}$cm. 
As we discussed in \S \ref{scalings},
$\sigma$ decreases as $\propto 1/z$
in the non-force-free regime $\sigma<$ a few
(and even faster when $\sigma>$ a few).

The assumption $B_z \gg B_\varpi$ -- required in order to derive the $z$ self-similar model --
does not allow us to examine the wind near the equator.
Nevertheless, 
the presented numerical solutions serve as the first examples of low-$\sigma$ asymptotic flows,
and, more importantly, they guided us to think which is the
situation in the most general (non-self-similar) case. Our main
conclusions (described and justified 
in \S~\ref{character} and \S \ref{genanalysis}) are in summary:
\begin{itemize}
\item
Type Ia conical flows (eq. [\ref{straightshape}]) are efficient accelerators.
The $\sigma$ decreases much faster than logarithmically.
\item
The same for the parabolic Ia, IIa (they likely apply to nonrelativistic flows).
\item
For sufficiently small asymptotic values of the $\sigma$ function,
a transition $\sigma=\sigma_{c}$ exist, where the azimuthal centrifugal force
is comparable to the electromagnetic force in the transfield direction.
The acceleration may or may not continue after this transition. In both
cases, however, the solvability condition at infinity 
(Heyvaerts \& Norman 1989; Chiueh et al. 1991)
is not mandatory. The latter should be satisfied only in cases where $\sigma_\infty > \sigma_{c}$.
Note that for the equatorial pulsar wind the ratio $f_{EM1}/f_{C \bot} \sim \sigma x^2 / \gamma^2 $
is $\gg 1$ at the position of the shock 
(for $\sigma \approx 0.003$, $x \sim 10^9$, and $\gamma \approx 10^6$).
However, the solvability condition may not be satisfied,
if the value of the $\sigma$ required at the shock is not the asymptotic $\sigma_\infty$, but the
$\sigma$ function still decreases at the position of the shock.
Moreover, in applications to jets related to other astrophysical settings, the transition
$\sigma=\sigma_{c}$ may be important.
\item
In flows where the acceleration continues to $\sigma < \sigma_{c}$,
the line shape is hyperbolic and the curvature radius of the poloidal field-streamlines
becomes negative.
\end{itemize}
A required condition for the above acceleration mechanism to work is 
$\mu \gg \sigma_{\rm M} $. As we discuss in \S~\ref{genanalysis},
this condition means that at the inner super-Alfv\'enic regime $B_p \varpi^2$ should be $\gg A $,
and the current that flows along the field lines should be much larger than the typical 
values obtained in monopole solutions, $| I | \gg A \Omega$.
We argue in \S~\ref{equatorialwind} that this condition is likely satisfied for the equatorial wind.
The scenario presented in that section is outlined below with the help of the figure \ref{fig7}.
The various transitions/regimes as we move along the reference (thicker)
field line, marked with the numbered vertical dotted lines are:
\begin{figure*}
\centering
  {\includegraphics[width=1.\textwidth]{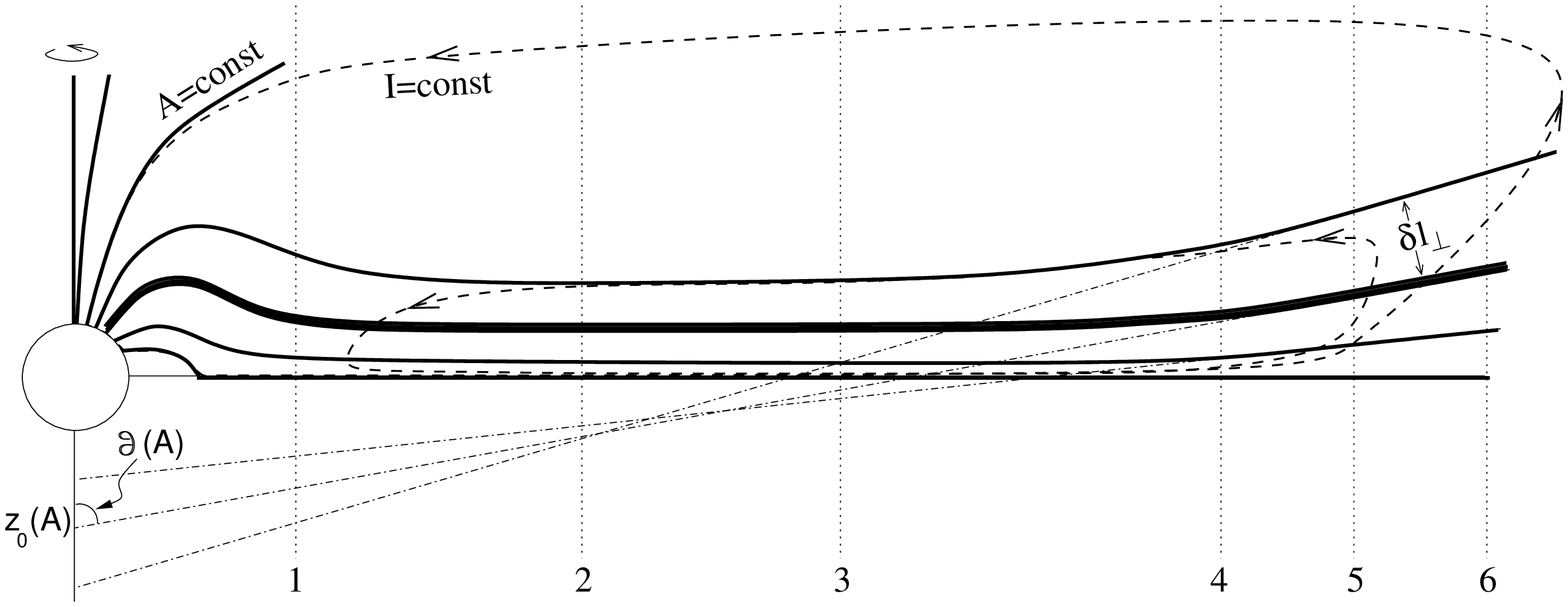}}
  \caption{
Sketch of an equatorial pulsar wind (not in scale). 
The lines of the poloidal magnetic field -- the $A=const$
bold lines --
have a dipolar shape close to the source and after a 
phase where they are roughly parallel to the equator
(bunched within a small solid angle), they become type conical Ia 
[$z=z_0(A) + \varpi / \tan \vartheta(A)$].
The lines of the poloidal current $\boldsymbol{J}_p$ -- the $I=const$
dashed lines -- are also shown.
The various transitions/regimes marked with the numbered vertical dotted lines are
discussed in \S~\ref{conclusions}.
\label{fig7}}
\end{figure*}
\smallskip \\ --
Origin $\rightarrow$ 1: 
This is the most uncertain regime, and very important problem for future research.
We can only speculate what may happen there.
The poloidal field possibly starts with a dipolar shape, however, inertial effects
(associated with the azimuthal velocity) create an open magnetosphere. 
(Note that even under force-free conditions the last closed field line does not
in principle intersect the equator at the light cylinder; see \citealp{U03}).
Non-ideal MHD effects accelerate created pairs to $\gamma_1 \gg 1$.
One has to solve the dynamics by including the back-reaction from the various emission mechanisms
to yield the exact value of $\gamma_1$. 
In principle, this value could correspond to 
matter-dominated flow $\sigma_1 \ll 1$ (although the poloidal field is strong, the Poynting flux
associated with the $E B_\phi$ product may not be larger than the matter energy flux).
The initial conditions after the pair acceleration in the non-ideal
MHD regime may correspond even to super-Alfv\'enic conditions in which case the
light cylinder has not a particular physical meaning.\footnote{
In cases where the flow starts with super-Alfv\'enic speed, the 
$\Omega $ is no longer related to the matter rotation. 
It is just the ratio $c E / \varpi B_p $ \citep[e.g.,][]{C95,VK03b}.}
\smallskip \\ --
1: Here $x \sim$ a few. 
This may be the base of the ideal MHD regime (although there should be non-ideal effects 
near the midplane even after that point, see \S~\ref{equatorialwind}).
Downstream from this point the super-Alfv\'enic asymptotic analysis holds.
According to equation (\ref{bpbphi}), 
$x B_p \approx (-B_\phi)(1-1/2 \gamma^2+1/2x^2) \approx -B_\phi$, and
$| I | = (c/2) \varpi | B_\phi | \approx (\Omega / 2) B_p \varpi^2$.
Also the poloidal curvature is $\varpi / {\cal R} \ll 1$ and the poloidal field lines 
are approximately straight and roughly parallel to the equator.\footnote{
Two other factors that could help the squeezing of the field lines
in a regime close to the equator are gravity and ram pressure
of infalling material from the environment of the pulsar.
However, both effects are probably insignificant.}
\smallskip \\ --
1 $\rightarrow$ 2: The product $B_p \varpi^2$ increases as 
$2 \pi \varpi^2 \delta A/ \delta S = \varpi \delta A / \delta l_\bot$, where
$\delta l_\bot$ is the distance between two neighboring field lines 
$A$, $A+\delta A$ (see fig. \ref{fig7}).
Since the $\varpi E \propto B_p \varpi^2$ increases, $| I | = (c/2) \varpi | B_\phi |$
should also increase in order to keep the inequality $B^2>E^2$ true.
Thus, matter kinetic energy (and possibly thermal energy) is transfered
to Poynting. 
Figure \ref{fig7} shows a current loop that
crosses the reference field line between points 1 and 2.
Here $J_\bot <0$ and the associated Lorentz force $(c/4 \pi) J_\bot B_\phi$ 
{\emph {decelerates}} the matter, increasing the $\sigma$ function.
Also $J_\parallel <0$, $J^0<0$, 
but the transfield component of the Lorentz force [$f_{B \bot} + f_{E \bot} \approx
(c/4 \pi) ( J^0 E + J_\parallel B_\phi)$] remains small, resulting in
$\varpi / {\cal R} = {\cal O}\left( {{\rm max} }\{1/x^2\,, 1/\gamma^2\}\right)$, 
\citep[e.g.][]{CLB98}.
\smallskip \\ --
2: Here the Lorentz factor is sufficiently small ($\lesssim 10$) 
for the field lines to start slightly bending. 
This is the base for the accelerating ideal MHD regime 
described in the previous sections.
This point corresponds to an ``initially'' Poynting flux-dominated wind.
The ``initial'' conditions satisfy 
$B_p \varpi^2 = (B_p \varpi^2)_i \gg A \Leftrightarrow \mu \gg \sigma_{\rm M}$
and the $\sigma$ function is $\gg 1$.
\smallskip \\ --
2 $\rightarrow$ 4: 
The field lines have a parabolic shape:
$\delta l_\bot / \varpi \delta A $ slightly increases,
resulting in a tiny decrease in $B_p \varpi^2$.
The $\sigma$ function decreases according to the equation (\ref{bernoullifinal}).
For the poloidal current density, $J_\parallel <0$, $J_\bot > 0$.
\smallskip \\ --
3: Classical fast-magnetosound point, $\gamma_3\approx\mu^{1/3}$, $\sigma_3 \approx \mu^{2/3}$.
\smallskip \\ --
4: The $\sigma$ function is unity. Equation (\ref{bernoullifinal}) implies that 
$B_p \varpi^2 =(B_p \varpi^2)_i/2$. 
\smallskip \\ --
4 $\rightarrow$ 6: This is the conical Ia phase 
(the discussion of \S~\ref{genanalysis} refers to this regime).
The field lines become $z=z_0(A) + \varpi / \tan \vartheta(A)$ 
(tangent to the dot-dashed lines in fig. \ref{fig7}).
It is seen in figure \ref{fig7} how the conditions $z_0^{'}>0$, $\vartheta^{'} > 0$ are realized.
The $B_p \varpi^2$ decreases according to equation (\ref{straightfield}). As a result,
the $\sigma$ function decreases from $\approx 1$
to $\sigma_{\rm {min}}$ given by equation (\ref{sigmamin}).
\smallskip \\ --
5: The current loop that the reference field line crosses at this point has $J_\parallel=0$, i.e.,
here the current-carrying regime ($J_\parallel<0$) ends and the return-current 
regime begins ($J_\parallel>0$).
\smallskip \\ --
6: Here the $B_p \varpi^2 /A$ is of the order of unity. The fast acceleration cannot continue after
that point. However, due to a tiny deviation from a $\varpi / {\cal R} = 0 $ it may continue 
logarithmically (see \citealp{CLB98} and references therein).

If at point 6 the value of $\sigma_6$ is $\sigma_{s}=0.003$
while at point 2 the flow is Poynting dominated ($\sigma_2 \gg 1$), then
the ratio $(B_p \varpi^2)_6/(B_p \varpi^2)_2 = \sigma_6 $ and the required 
value for the $B_p \varpi^2/A$ at point 2 is $1/ \sigma_6 \approx 300$.
Assuming further that ideal MHD holds during the 1 $\rightarrow$ 2 phase
and that the flow started at 1 with $\sigma_1\ll 1$, we get
$(B_p \varpi^2)_2/(B_p \varpi^2)_1 =
(B_\phi \varpi)_2/(B_\phi \varpi)_1 = 1/\sigma_1$.
Thus, $(B_p \varpi^2)_1/A = \sigma_1/ \sigma_6 \approx 300 \sigma_1$.
For a dipolar field before the point 1, $(B_p \varpi^2)_1/A =2$, and we get
a required value for $\sigma_1 = 2 \sigma_6 \approx 0.006$.
However, it is not clear that all the above assumptions
hold, so we cannot be sure for the derived value $\sigma_1$. 

In general, the fast (faster than logarithmic) acceleration
in magnetized outflows gives asymptotically 
$\sigma_{\rm min} \sim A/ B_p \varpi^2 = \sigma_{\rm M}/\mu$,
where $B_p \varpi^2$ is the ``initial'' value at the Poynting dominated regime
(point 2 in fig. \ref{fig7}).
Besides the application to pulsar winds on which we focussed in this paper,
the formulation presented is quite general and AGN or GRB relativistic
as well as YSO nonrelativistic jet asymptotics can also be considered
(the asymptotic analysis presented in \S~\ref{MHD} applies 
in all these cases, no matter if asymptotically $\sigma \ll 1$ or $\sim 1$, or
even $> 1$).
In addition, the $z$ self-similar model of \S~\ref{zss} could be used for examining
AGN or GRB jets in their super-Alfv\'enic regime.

\acknowledgements 
The author is grateful for useful discussions with Arieh K\"onigl and Ruben Krasnopolsky.
This work was supported in part by NASA grant
NAG5-12635 and by the U.S. Department of Energy under Grant No. B341495
to the Center for Astrophysical Thermonuclear Flashes at the University of Chicago.
Support from a McCormick Fellowship at the Enrico Fermi Institute is also acknowledged.

\end{document}